\documentclass[aps,prd,preprintnumbers,superscriptaddress,nofootinbib,11pt]{revtex4}%
\usepackage[dvipdfmx]{graphicx}
\usepackage{bm,latexsym,amsmath,amssymb,amsfonts,mathrsfs}
\input{colordvi.tex}
\usepackage {cancel}
\usepackage{hyperref}
\usepackage{url}
\hypersetup{
    colorlinks=true,
    citecolor=cyan,
}
\newcommand*{\D}{{\rm d}}
\newcommand*{\mpl}{M_{\rm Pl}}

\begin{document}

\title{
Gravitational mode mixing around black holes in scalar-tensor theories with parity-violating terms
}

\author{Shin'ichi Hirano}
\affiliation{National Institute of Technology, Oyama College
771 Nakakuki, Oyama-shi, Tochigi 323-0806, Japan}
\affiliation{Department of Physics, Rikkyo University, Toshima, Tokyo 171-8501, Japan}

\author{Masashi Kimura}
\affiliation{Department of Information, Artificial Intelligence and Data Science,
Daiichi Institute of Technology, Tokyo 110-0005, Japan}
\affiliation{Department of Physics, Rikkyo University, Toshima, Tokyo 171-8501, Japan}

\author{Masahide Yamaguchi}
\affiliation{Cosmology, Gravity and Astroparticle Physics Group, Center for Theoretical Physics of the Universe, Institute for Basic Science, Daejeon 34126, Korea}
\affiliation{Department of Physics, Tokyo Institute of Technology,
2-12-1 Ookayama, Meguro-ku, Tokyo 152-8551, Japan}

\preprint{RUP-24-5}

\begin{abstract}
We investigate black holes and gravitational perturbations when both the scalar Gauss-Bonnet and dynamical Chern-Simons gravity sectors coexist in addition to the Einstein-Hilbert term, and both sectors are coupled to a single canonically normalized scalar field.
The presence of the scalar Gauss-Bonnet gravity sector allows the scalar field to possess a non-vanishing background solution, resulting in additional couplings between odd- and even-type gravitational perturbations arising from the dynamical Chern-Simons gravity sector. 
We illustrate the impact of these even-odd gravitational couplings in gravitational perturbations around a static spherically symmetric black hole.
Although the couplings between the odd and even-type gravitational perturbations are known to appear in purely tensorial gravity theories with higher-curvature corrections, we demonstrate it in scalar-tensor theories.
\end{abstract}
\maketitle

\section{Introduction}\label{sec: introduction}

Since the first detection of gravitational waves (GWs) from a black hole binary merger by the LIGO/Virgo experiment~\cite{LIGOScientific:2016aoc,LIGOScientific:2016vbw}, numerous GW events have been observed by interferometers. 
GWs emitted from binary black holes contain valuable information in the strong-gravity regime and provide an opportunity to test theories of gravity.
The observed waveforms closely match those predicted by general relativity (GR)~\cite{LIGOScientific:2016aoc}.
Future precise observations could detect deviations from general relativity through gravitational waves (GWs) if physics beyond general relativity exists. 
During the ringdown phase of a black hole merger, GWs are computed as perturbations around the final remnant black hole. 
Quasinormal modes (QNMs) represent the characteristic frequencies of perturbations around a black hole, and the GWs during the ringdown phase are described by a superposition of QNMs (for reviews~\cite{Nakamura:1987zz,Kokkotas:1999bd,Nollert:1999ji,Ferrari:2007dd,Berti:2009kk,Konoplya:2011qq,Hatsuda:2021gtn}).

General relativity (GR) faces several challenges of the late-time acceleration~\cite{SupernovaSearchTeam:1998fmf,SupernovaCosmologyProject:1998vns}, the formation of black hole singularities~\cite{Penrose:1964wq,Hawking:1970zqf,Hawking:1976ra}, and the quantization of gravity at both low- and high-energy scales.
These strongly suggest that the theory of gravity should be modified from general relativity.
In most cases, the modification of gravity involves adding extra degrees of freedom to the metric.
In fact, the simplest modification is to introduce a non-minimal coupling of the metric to a scalar field, {\it i.e.}, a scalar-tensor theory.
As a typical example, we focus both on the scalar Gauss-Bonnet (sGB) gravity~\cite{Metsaev:1987zx,Kanti:1995vq,Green:1984sg,Antoniadis:1992sa} and the dynamical Chern-Simons (dCS) gravity~\cite{Lue:1998mq,Jackiw:2003pm,Alexander:2009tp}.
Both theories of gravity are known as low-energy theories of string theory~\cite{Mignemi:1992nt, Antoniadis:1993jc, Mignemi:1993ce, Kanti:1995vq,Daniel:2024lev}.
The sGB gravity sector breaks the four-dimensional topological invariance of the Gauss-Bonnet term by coupling the dynamical scalar field, whereas the dCS gravity sector has a parity-violating coupling between the scalar field and the Pontryagin term.\footnote{We note that the sGB gravity is a healthy classical theory included in the Horndeski theory which is the most general theory with the second-order equations of motion \cite{Horndeski:1974wa,Deffayet:2011gz,Kobayashi:2011nu}.} 
The dCS gravity should be regarded as a low-energy effective field theory because of the presence of ghost modes around a static spherically symmetric black hole~\cite{Motohashi:2011ds} and any spacetime~\cite{Crisostomi:2017ugk}.
The black hole solutions and QNMs have been studied in the sGB gravity~\cite{Yunes:2011we,Blazquez-Salcedo:2016enn,Blazquez-Salcedo:2017txk,Blazquez-Salcedo:2018pxo,Konoplya:2019hml,Zinhailo:2019rwd,Churilova:2019sah,Blazquez-Salcedo:2020caw, Pierini:2021jxd,Pierini:2022eim} and the dCS gravity~\cite{Yunes:2007ss,Cardoso:2009pk,Molina:2010fb,Yunes:2011we,Srivastava:2021imr,Wagle:2021tam,Blazquez-Salcedo:2023hwg}, respectively, where the deviations from the GR case can be expressed as a series of small couplings.
Recently, in the context of the effective field theory (EFT), corrections to black hole solutions and QNMs with small EFT couplings have been studied~\cite{Endlich:2017tqa,Cardoso:2018ptl,Cano:2019ore,deRham:2020ejn,Nomura:2021efi,Cano:2022wwo,Silva:2022srr,Cano:2023jbk,Cayuso:2023xbc,Cano:2023tmv,Melville:2024zjq}.

In this paper, we study black holes and gravitational perturbations when both the sGB and dCS gravity sectors coexist, and both sectors are coupled to a single scalar field.
The presence of the sGB gravity sector allows the scalar field to have a background value, leading to additional new couplings between the odd and even-type gravitational perturbations appearing through the dCS gravity sector.
We illustrate the impact of the even-odd gravitational couplings in the perturbations around the static spherically symmetric black hole solutions.
While the couplings between the odd- and even-type gravitational perturbations are known to appear in purely tensorial gravity theories with higher-curvature corrections~\cite{Cardoso:2018ptl,deRham:2020ejn,Cano:2021myl,Cano:2023jbk}, we demonstrate it in scalar-tensor theories.

The paper is structured as follows: in Sec.~\ref{sec: theory}, we introduce a system in which the scalar Gauss-Bonnet gravity sector and the dynamical Chern-Simons gravity sector coexist. 
We then construct static spherically symmetric black hole solutions. In Sec.~\ref{sec: perturbation}, we present the master equations for gravitational perturbations around the static spherically symmetric black hole solutions. 
In Sec.~\ref{sec: QNM}, we analyze the quasinormal frequency for each mode. 
In Sec.~\ref{sec: rotating BH}, we discuss slowly rotating black hole solutions without the $\mathbb{Z}_2$ symmetry. 
In Sec.~\ref{sec: tidal}, we discuss the tidal response due to the even-odd gravitational couplings. 
Finally, we summarize our results in Sec.~\ref{sec: summary}.
We use the natural unit, $c=\hbar=1$.

\section{Equations of motion and static spherically symmetric black hole solutions}\label{sec: theory}

\subsection{Scalar Gauss-Bonnet and dynamical Chern-Simons system}\label{sec: action}

We consider gravity theories with both the sGB sector~\cite{Metsaev:1987zx,Kanti:1995vq,Green:1984sg,Antoniadis:1992sa} and dCS sector~\cite{Lue:1998mq,Jackiw:2003pm,Alexander:2009tp} in 4 dimensions,\footnote{We use the notation of the sGB coupling in Ref.~\cite{Nakashi:2020phm} while that of the dCS coupling is different from Ref.~\cite{Cardoso:2009pk} by the numerical factor $8$.}
\begin{align}
   {\cal L}
    &= \frac{\mpl^2}{2}R -\frac{1}{2}(\partial\phi)^2 
    -\frac{1}{2}m^2\phi^2
    +\frac{b_1}{\Lambda} \phi(R^2 -4R_{\mu\nu}R^{\mu\nu} +R^{\mu\nu}{}_{\rho\sigma}R^{\rho\sigma}{}_{\mu\nu})
    +\frac{b_2}{\Lambda}\phi \tilde{R}^{\mu\nu}{}_{\rho\sigma}R_{\nu\mu}{}^{\rho\sigma},
    \label{eq:lagrangian}
\end{align}
where $\mpl$ is the Planck mass, $\Lambda$ is the cutoff scale, 
$b_1$ and $b_2$ are constants,
${\tilde R}^{\mu\nu}_{\quad\rho\sigma} =\varepsilon^{\mu\nu\alpha\beta}R_{\alpha\beta\rho\sigma}$, $\varepsilon^{\mu\nu\rho\sigma} = \epsilon^{\mu\nu\rho\sigma}/\sqrt{-g}$, and $\epsilon^{\mu\nu\rho\sigma}$ is an anti-symmetric symbol with $\epsilon^{0123} = 1$.
The first three terms represent the free part of the fields, while the fourth and fifth terms represent the sGB coupling and the dCS coupling, respectively.
A similar setup has been investigated in cosmology~\cite{Satoh:2007gn}. 
In this paper, we consider the weak couplings of a single scalar field and curvatures of spacetime.
The equations of motion and their solutions can be obtained by considering the zeroth-order solutions and the corrections written as a series of small coupling terms.
The zeroth-order solutions are determined by the free part, and according to the no-hair theorem~\cite{Bekenstein:1971hc,Bekenstein:1995un}, these solutions coincide with the vacuum solution of GR and exhibit Ricci flatness.
We consider leading relevant scalar-tensor couplings in black holes.
Using conformal transformations, the quadratic derivative terms with the scalar field and curvatures, such as $\phi R$, can be transformed into canonical free parts and higher-order self-couplings for the fields.
At the next order of derivative terms, terms such as $\phi R^2$ and $\phi R_{\mu\nu}^2$ vanish in the equations of motion due to the Ricci flatness. 
The leading-order scalar-tensor couplings are the ``linear" sGB and dCS couplings as depicted in Eq.~\eqref{eq:lagrangian}.
In the context of the EFT extension of GR 
with a single scalar field
around black holes, the sGB and dCS couplings in Eq.~\eqref{eq:lagrangian} are also leading operators~\cite{ours}.\footnote{
As shown later, the non-GR effect in the QNMs of the black holes in Eq.~\eqref{eq:lagrangian} emerges at ${\cal O}(1/\Lambda^2)$.
While the $\phi^2$ coupling to the quadratic curvature term $\phi^2R_{\mu\nu\rho\sigma}^2/\Lambda^2$ 
affects the scalar QNM spectra at ${\cal O}(1/\Lambda^2)$, such terms do not have an impact on the gravitational QNM spectra at the same order.
In this paper, we focus primarily on the gravitational QNMs to demonstrate the gravitational mixing effect in scalar-tensor theories. Therefore, for simplicity, we do not consider these terms.
}
The equation of motion for $\phi$ is given by
\begin{align}
    &(\Box - m^2)\phi 
    = -\frac{b_1}{\Lambda}
    (R^2 -4R_{\mu\nu}R^{\mu\nu} +R^{\mu\nu}{}_{\rho\sigma}R_{\mu\nu}{}^{\rho\sigma})
    -\frac{b_2}{\Lambda}\tilde{R}^{\mu\nu}{}_{\rho\sigma}R_{\nu\mu}{}^{\rho\sigma}.
    \label{eq: scalarEoM}
\end{align}
Varying the action with respect to the metric $g_{\mu\nu}$, we also obtain the equations of motion for the metric
\begin{align}
    \frac{\mpl^2}{2}\left(R_{\mu\nu}-\frac{1}{2}g_{\mu\nu}R\right)
    &=
    \frac{1}{2}\partial_{\mu}\phi\partial_{\nu}\phi
    -\frac{1}{2}g_{\mu\nu}\left\{\frac{1}{2}(\partial\phi)^2 +\frac{1}{2}m^2\phi^2\right\}
  \notag\\    
    &\quad
    -\frac{b_1}{\Lambda} \Bigr\{
    -2R\nabla_{\mu}\nabla_{\nu}\phi  +2(g_{\mu\nu}R-2R_{\mu\nu})\Box\phi
    \notag\\
    &\hspace{1.55cm}
    +8R_{\rho(\mu}\nabla^{\rho}\nabla_{\nu)}\phi
    -4g_{\mu\nu}R^{\rho\sigma}\nabla_{\rho}\nabla_{\sigma}\phi
    +4R_{(\mu |\rho|\nu) \sigma}\nabla^{\rho}\nabla^{\sigma}\phi
    \Bigr\}
\notag\\
    &\quad -\frac{8b_2}{\Lambda}\Bigl\{
    \nabla_{\rho}\phi \,\varepsilon^{\rho\sigma}{}_{\alpha(\mu}\nabla^{\alpha}R_{\nu)\sigma}
    +\frac{1}{2}\nabla^{\rho}\nabla^{\sigma}\phi \tilde{R}_{\rho(\mu\nu)\sigma}
    \Bigr\}.
    \label{eq: gravityEoM}
\end{align}
Hereafter, for simplicity, we consider a massless scalar field.

\subsection{Static spherically symmetric black hole solutions}\label{sec: static BH}

Let us consider static spherically symmetric black hole solutions.
The ansatz for the static spherically symmetric metric and the scalar field is as follows
\begin{align}
     ds^2 &= \bar{g}_{\mu\nu}dx^{\mu}dx^{\nu} 
    = -A(r)dt^2 +\frac{1}{B(r)}dr^2 +r^2(d\theta^2 +\sin{\theta}^2d\varphi^2),
    \label{eq:gbar}
\\
    \phi &= \bar{\phi}(r).
    \label{eq:phibar}
\end{align}
The sGB and dCS couplings affect the solutions as corrections.
The zeroth-order solutions are given by the GR ones, {\it i.e.}, the Schwarzschild metric.
The corrections are characterized by the dimensionless small parameter
\begin{align}
    \epsilon := \frac{1}{\Lambda\mpl r_{\rm g}^2},\label{eq: EFT parameter}
\end{align}
where $r_{\rm g}$ is the Schwarzschild radius.
As discussed below, the leading correction to the scalar field is of the order of ${\cal O}(\epsilon)$ whereas the correction to the metric is of the order of ${\cal O}(\epsilon^2)$.
In the following section, the equations of motion are solved up to ${\cal O}(\epsilon^2)$ to derive the corrections to the GR solutions.

Due to the no-hair theorem in GR~\cite{Bekenstein:1971hc,Bekenstein:1995un}, there is no scalar field configuration and no modification to the Schwarzschild metric at ${\cal O}(\epsilon^{0})$.\footnote{
The massless scalar field can be constant as the lowest-order solution, but such a solution makes no additional contribution to the higher-order equations of motion. In this paper, $\bar{\phi}={\rm const.}$ brunch is not considered.}
We set the metric and scalar field to
\begin{align}
    A(r) &= 1 -\frac{r_{\rm g}}{r} +\epsilon\, a_{\rm 1st}(r) +\epsilon^2 a_{\rm 2nd}(r),
    \label{eq:agbar}
    \\
    B(r) &= 1 -\frac{r_{\rm g}}{r} +\epsilon\, b_{\rm 1st}(r) +\epsilon^2b_{\rm 2nd}(r),\\
    \bar{\phi}(r) &= \epsilon\, \pi_{\rm 1st}(r) +\epsilon^2\pi_{\rm 2nd}(r).
\end{align}
Substituting this ansatz into the equations of motion, Eqs.~\eqref{eq: scalarEoM} and~\eqref{eq: gravityEoM}, and solving perturbatively in $\epsilon$, one can determine the unknown functions, $a$, $b$, and $\pi$. 
Its structure is as follows. 
At the lowest order, the equations of motion are automatically satisfied since they are solutions of GR.
At first order, the correction of the scalar field $\pi_{\rm 1st}$ is sourced by the GR solution on the right-hand side of Eq.~\eqref{eq: scalarEoM} while those for the metric, $a_{\rm 1st}$ and $b_{\rm 1st}$, are not sourced by the right-hand side of Eq.~\eqref{eq: gravityEoM} due to the vanishing scalar field in GR.
On the other hand, at second order, the correction of the scalar field $\pi_{\rm 2nd}$ is not sourced by the right-hand side of Eq.~\eqref{eq: scalarEoM} due to the absence of first-order solutions for the metric while the corrections for the metric, $a_{\rm 2nd}$ and $b_{\rm 2nd}$, are sourced by the right-hand side of Eq.~\eqref{eq: gravityEoM}.
The unknown functions are determined analytically,
\begin{align}
    a_{\rm 1st}(r) &=b_{\rm 1st}(r)=0,\label{eq: ab1st}
\\
     \frac{\pi_{\rm 1st}(r)}{\mpl} &= 4b_1\frac{r_{\rm g}}{r} \left(1+\frac{r_{\rm g}}{2r} +\frac{r_{\rm g}^2}{3r^2}\right),\label{eq: piBG 1st}
\\
    a_{\rm 2nd}(r) &= \frac{4b_1^2}{3}\frac{r_{\rm g}^3}{r^3}\left(1 +\frac{13r_{\rm g}}{r} +\frac{33r_{\rm g}^2}{10r^2} +\frac{12r_{\rm g}^3}{5r^3} -\frac{5r_{\rm g}^4}{r^4}\right),\label{eq: a2nd}
\\
    b_{\rm 2nd}(r) &= 8b_1^2\frac{r_{\rm g}^2}{r^2}\left(1 +\frac{r_{\rm g}}{2r} +\frac{13r_{\rm g}^2}{3r^2} +\frac{r_{\rm g}^3}{4r^3} +\frac{r_{\rm g}^4}{5r^4} -\frac{23r_{\rm g}^5}{6r^5}\right),\label{eq: b2nd}
\\
    \pi_{\rm 2nd}(r) &=0.\label{eq: pi2nd}
\end{align}
We have required that the solutions are regular at $r=r_{\rm g}$.
Under parity transformation, the static spherically symmetric metric and scalar field have even parity. 
Therefore, the dCS term that breaks parity symmetry does not affect the static spherically symmetric solutions, which is exactly the reason why these background solutions depend only on $b_1$.
Our result aligns with this symmetry argument and reproduces the previous results in the weak coupling regime~\cite{Yunes:2011we,Blazquez-Salcedo:2016enn}. 
Due to the corrections in the metric, the horizon scale deviates from the Schwarzschild radius.
From the condition $A(r_{\rm H})= {\cal O}(\epsilon^3)$, the horizon radius $r_{\rm H}$ becomes
\begin{align}
    r_{\rm H} = r_{\rm g}\left(1 -\epsilon^2b_1^2\frac{98}{5}\right).
\end{align}
The thermodynamics are studied in the context of black holes in the Horndeski theory~\cite{Minamitsuji:2023nvh}.\footnote{
For fixed $\epsilon, b_1$ and $M$, the black hole solution in our setup is uniquely determined.
This implies that the solution is thermodynamically stable, which is consistent with
the results in Sec.~\ref{sec: QNM} where the quasinormal modes correspond to damped oscillations.}

\section{Gravitational Perturbations}\label{sec: perturbation}

In this section, we study gravitational perturbations around the static spherically symmetric background solutions $\bar{g}_{\mu\nu}$ and $\bar{\phi}$ in Eqs.~\eqref{eq:gbar} and \eqref{eq:phibar} with Eqs.~\eqref{eq:agbar}--\eqref{eq: pi2nd}.
The perturbed scalar field is given by
\begin{align}
    \phi(t,r,\theta) = \bar{\phi}(r) +e^{-i\omega t}\Pi(r)Y_{\ell 0}(\theta).\label{eq: BGscalar+pert}
\end{align}
The time evolution is assumed to be characterized by the factor $e^{-i\omega t}$, while the angular dependence is represented by the spherical harmonics $Y_{\ell 0}$ with the azimuthal number  $m=0$.\footnote{
Because of the spherical symmetry, it is enough to consider the $m = 0$ case.
Other spherical harmonics with $m \neq 0$ can be obtained by acting the ladder operators constructed from the generator of the spherical symmetry.
}
We mainly focus on $\ell \geq 2$ modes.
The metric perturbation can be decomposed by parity symmetry in the case of GR. At zeroth order in $\epsilon$, the theory is GR, and then we can take the standard Regge-Wheeler gauge~\cite{Regge:1957td} for the metric perturbations.
The first-order corrections are sourced by the GR solution.
So, higher-order corrections are also determined by the GR solution.
We still use the background metric and the standard metric perturbations with the Regge-Wheeler gauge,
\begin{align}
    g_{\mu\nu} &= \bar{g}_{\mu\nu} +h_{\mu\nu}^{-} +h_{\mu\nu}^{+},\label{eq: BGmetric+pert}
\\
    h_{\mu\nu}^{-}dx^\mu dx^\nu &= 2e^{-i\omega t}\sin{\theta}\partial_{\theta}Y_{\ell0}d\phi(h_0 dt+h_1dr),
\\
    h_{\mu\nu}^{+}dx^\mu dx^\nu &=
    e^{-i\omega t}Y_{\ell0}\bigg\{AH_0dt^2+2H_1dtdr
    +\frac{H_2}{B}dr^2 +r^2K(d\theta^2+\sin^2{\theta}d\phi^2)\bigg\},
\end{align}
where $h_0,~h_1,H_0,~H_1,~H_2$, and $K$ are functions of $r$.
The odd ($-$) mode components $h_0,~h_1$ and the even ($+$) mode components $H_0,~H_1,~H_2, K$ are not independent, but related through the equations of motion. 
In our setup, the metric perturbations have two helicity-2 modes.

\subsection{Master equations}\label{sec: Master Eqs}

To derive the master equations for the scalar field, odd and even-type gravitational modes, we substitute Eqs.~\eqref{eq: BGscalar+pert} and~\eqref{eq: BGmetric+pert} into the equations of motion, Eqs.~\eqref{eq: scalarEoM} and~\eqref{eq: gravityEoM}, and expand the equations up to ${\cal O}(\Pi)$, ${\cal O}(h_{\mu\nu})$, and ${\cal O}(\epsilon^2)$.
At the lowest order, $\epsilon=0$, the master equations are decoupled and given by the single master variables as in the case of GR,
\begin{align}
    \Psi^{\rm s} &= \frac{r\Pi}{\mpl},\label{eq: master var scalar GR}
\\
    \Psi^{-} &= \left(1-\frac{r_{\rm g}}{r}\right)\frac{i h_1}{r\omega},\label{eq: master var - GR}
\\
    \Psi^{+} &= \frac{1}{\lambda r-3r_{\rm g}}\left\{-r^2K+\left(1-\frac{r_{\rm g}}{r}\right)\frac{irH_1}{\omega}\right\},\label{eq: master var + GR}
\end{align}
where $\lambda = \ell^2+\ell-2$. The master equations  are given by
\begin{align}
   F_{\rm GR}(r)\frac{\D}{\D r}\left(F_{\rm GR}(r)\frac{\D \Psi^i}{\D r}\right) +\left\{\omega^2 -F_{\rm GR}(r)V^i_{\rm GR}\right\}\Psi^i =0,
   ~(i={\rm s},-,+).
\end{align}
where $F_{\rm GR}(r) = 1-r_{\rm g}/r$, and
\begin{align}
    V^{\rm s}_{\rm GR} &= \frac{(\lambda +2)r +r_{\rm g}}{r^3},\label{eq: KG potential}
\\
    V^{\rm -}_{\rm GR} &= \frac{(\lambda +2)r -3r_{\rm g}}{r^3},\label{eq: RW potential}
\\
    V^{\rm +}_{\rm GR} &= \frac{1}{(\lambda r+3r_{\rm g})^2r^3}\bigl\{9r_{\rm g}^3 +9\lambda r_{\rm g}^2r +3\lambda^2r_{\rm g}r^2
    +\lambda^2(\lambda+2)r^3\bigr\}.\label{eq: Z potential}
\end{align}
The equations for the odd and even gravitational modes are called the Regge-Wheeler and Zerilli equations~\cite{Regge:1957td,Zerilli:1970se}.
At ${\cal O}(\epsilon)$, the scalar-odd mode and scalar-even mode couplings can appear on the right-hand side in the equations of motion, Eqs.~\eqref{eq: scalarEoM} and~\eqref{eq: gravityEoM}.
At ${\cal O}(\epsilon^2)$, the kinetic terms and potential terms are modified due to modification of the background metric. As in the case of the background solutions, there is no mixing among the scalar field and metric perturbations at second order.
However, new couplings can appear between the odd and even gravitational modes because the source terms from the dCS coupling do not vanish thanks to the background value of the scalar field.
The resultant master equations are given by
\begin{align}
     F(r)\frac{\D}{\D r}\left(F(r)\frac{\D {\bm \Psi}}{\D r}\right)+\left\{\omega^2({\bm I}+\Delta{\bm c^2}) -F(r)({\bm V}_{\rm GR} +{\bm V}_{\rm correction})\right\}{\bm \Psi} =0,\label{eq: coupled master eq}
\end{align}
where $F(r):=\sqrt{A(r)B(r)}$, and 
\begin{align}
    {\bm \Psi} &= 
    \begin{pmatrix}\tilde{\Psi}^{\rm s}\\
    \tilde{\Psi}^{-}\\
    \tilde{\Psi}^{+}
    \end{pmatrix},
\\
    \Delta{\bm c^2} &= \epsilon^2F(r)
    \begin{pmatrix}
    \Delta c_{\rm s}^2 & 0 & 0 \\
    0 & \Delta c_{-}^2 & 0 \\
    0 & 0 &\Delta c_{+}^2
    \end{pmatrix}\label{eq: sound speed},
\\
    {\bm V}_{\rm GR} &= 
    \begin{pmatrix}
    V^{\rm s}_{\rm GR}  & 0 & 0 \\
    0 & V^{-}_{\rm GR} &0 \\
    0 & 0 & V^{+}_{\rm GR} 
    \end{pmatrix}\label{eq: potential_GR},
\\
    {\bm V}_{\rm correction} &= 
    \epsilon
    \begin{pmatrix}
    0 & V^{\rm s-}_{\rm 1st} & V^{\rm s+}_{\rm 1st} \\
    V^{\rm s-}_{\rm 1st} &  0 & 0 \\
    V^{\rm s+}_{\rm 1st} & 0 & 0
    \end{pmatrix} 
    + \epsilon^2
    \begin{pmatrix}
    V^{\rm s}_{\rm 2nd} & 0 & 0 \\
    0 &  V^{\rm -}_{\rm 2nd} & V^{\rm -+}_{\rm 2nd} \\
    0 &  V^{\rm -+}_{\rm 2nd} & V^{\rm +}_{\rm 2nd}
    \end{pmatrix}\label{eq: potential_correction}.
\end{align}
The explicit forms of ${\bm \Psi}$, $\Delta {\bm c}^2$,
and ${\bm V}_{\rm correction}$ are written in App~\ref{sec: explicit forms of master eqs}.
Since $\Delta {\bm c}^2$ vanishes at the horizon, the propagation speed of each mode coincides with the speed of light.
This implies that information inside the horizon cannot propagate outside of the horizon.
The non-diagonal components of the potential matrix in Eq.~\eqref{eq: potential_correction} represent the mode mixing.
We note that the scalar-odd and scalar-even coupling terms at ${\cal O}(\epsilon^2)$ in Eq.~\eqref{eq: potential_correction} vanish because the next leading terms appear at ${\cal O}(\epsilon^3)$.
Considering both sGB and dCS couplings, new terms, $V^{-+}_{\rm 2nd}$, can appear in the master equations. 
In particular, the even-odd gravitational couplings have been known to be a generic feature of purely tensorial gravity theories with higher-curvature corrections~\cite{Cardoso:2018ptl,deRham:2020ejn,Cano:2021myl,Cano:2023jbk}.
We find that this is characteristic not only in purely tensorial gravity theories but also when both the sGB term and the dCS term are present.
In the context of cosmology, it has been known that the even-odd gravitational couplings can appear in a similar system where both sGB and dCS gravity sectors coexist~\cite{Satoh:2007gn}.

The new even-odd gravitational couplings are of the order of $\epsilon^2$ in Eq. \eqref{eq: potential_correction} while the scalar-odd and scalar-even couplings are of the order of $\epsilon^1$.
At first glance, the latter couplings lead the observables.
However, the effect of the new couplings and that of the scalar couplings give the same order contributions to the QNMs.
With the new even-odd gravitational couplings, our model has new features in slowly rotating solutions and the tidal response sourced by an external gravitational object.

\section{Quasinormal Modes}\label{sec: QNM}

QNMs are the characteristic oscillation modes of perturbations around black holes.
The quasinormal frequencies are defined by the solutions of $\omega$ in the master equations imposing the purely ingoing condition at the horizon and the purely outgoing condition at infinity.\footnote{In the context of initial value problems, QNMs are defined as poles in Green's function and they correspond to the purely ingoing condition at the horizon and the purely outgoing condition at infinity for massless fields in asymptotically flat black holes~\cite{Leaver:1986gd,Nollert:1992ifk,Andersson:1995zk,Andersson:1996cm,Kokkotas:1999bd,Nollert:1999ji,Berti:2009kk,Konoplya:2011qq}.}
The Leaver method~\cite{Leaver:1985ax} is used to calculate the quasinormal frequencies. This calculation method is briefly presented for both decoupled and coupled systems.
In the decoupled system, the master equation is described by a single master variable $\Psi$,
\begin{align}
   f(r)\frac{\D}{\D r}\left(f(r)\frac{\D \Psi}{\D r}\right) +\left\{\omega^2 -f(r)V(r)\right\}\Psi\label{eq: decoupled master eq}
=0,
\end{align}
where $V(r)$ is a potential term, and $f(r) = 1-r_{\rm H}/r$. 
The wave function $\Psi$ can be factorized by a series of $f$,
\begin{align}
    \Psi = e^{i\omega r_\ast}f^{-2i\omega r_{\rm H}} \sum^{\infty}_{j=0} a_{j}f^{j},\label{eq: decoupled Psi}
\end{align}
where $\D r/\D r_\ast = f$.  
The wave function $\Psi$ of this form satisfies
the QNM boundary condition at the horizon thanks to the prefactor $e^{i\omega r_\ast}f^{-2i\omega r_{\rm H}}$. 
Substituting this ansatz into the master equation~\eqref{eq: decoupled master eq} yields the recursion relation for the coefficients $a_{j}$.
For a given $a_0$, the higher-order coefficients $a_j~(j=1,~2,~\cdots)$ can be determined by the recursion relation\footnote{
Since the boundary condition at the horizon has already been imposed, the first term $a_0$ determines the solution of the master equation.
}
in the form
$a_{j} = G_j(\omega) a_0$, where $G_j(\omega)$ is a function in $j$ and $\omega$.
If the wave function~\eqref{eq: decoupled Psi} satisfies the QNM boundary condition at infinity $f = 1$, then the sum of the series $\sum^{\infty}_{j=0} a_{j}$ should converge.
Such a solution can be derived approximately by imposing $a_{j}=0$ for sufficiently large $j$.
Thus, we need to solve $G_j(\omega)=0$ for sufficiently large $j$, and the quasinormal frequencies are given by those solutions~\cite{Leaver:1985ax}.
In GR, for $\ell=2$ and the fundamental modes, the quasinormal frequency for the scalar field is $(0.96728774 -0.19359239i)/(2M)$~\cite{Berti:2009kk} where $M$ is the black hole mass, and for the odd and even gravitational modes, it is $(0.74734337 -0.17792463i)/(2M)$~\cite{Chandrasekhar:1975zza,Berti:2009kk}.
In the case of non-GR gravity theories, if the master equation is expressed as a perturbation from the GR case, the quasinormal frequencies can be obtained as small deviations from the GR values~\cite{Cardoso:2018ptl,Cardoso:2019mqo,McManus:2019ulj,Cano:2019ore,deRham:2020ejn,Kimura:2020mrh,Nomura:2021efi,Cano:2022wwo,Franchini:2022axs,Cano:2023jbk,Cayuso:2023xbc, Cano:2023tmv,Hatsuda:2023geo}.

In a coupled system, the master equations have non-diagonal parts of the potential matrix\footnote{
Changing the master variable, the master equations~\eqref{eq: coupled master eq}
can be written in the form of Eq.~\eqref{eq: decoupled master eq} (see App.~A in~\cite{Cardoso:2019mqo}).}
\begin{align}
   f(r)\frac{\D}{\D r}\left(f(r)\frac{\D {\bf\Psi}}{\D r}\right) +\left\{\omega^2{\bf I} -f(r){\bf V}(r)\right\}{\bf \Psi}\label{eq: decoupled master eq}
=0,
\end{align}
where
${\bf \Psi}$ is a multi-component wave function, ${\bf V}$ is a potential matrix, and ${\bf I}$ is a unit matrix.
The ansatz of the wave function which satisfies the QNM boundary condition at the horizon is generalized to
\begin{align}
    {\bf \Psi} = e^{i\omega r_\ast}f^{-2i\omega r_{\rm H}} \sum^{\infty}_{j=0} {\bf a}_{j}f^{j},
    \label{eq:multicomponentleaver}
\end{align}
where the coefficient ${\bf a}_j$ are also multicomponents.
Substituting this ansatz into the master equations also yields recursion relations for ${\bf a}_j$.  
Using the recursion relations, we can write ${\bf a}_j$ in the form ${\bf a}_j = {\bf M}_j(\omega)~{\bf a}_0$, where ${\bf M}_j(\omega)$ is a matrix of functions of $j$ and $\omega$.
The QNM boundary condition at the spatial infinity
corresponds to the convergence of the sum of the series in Eq.~\eqref{eq:multicomponentleaver} at infinity $f = 1$.
This implies that the limit of ${\bf a}_j$ vanishes as $j \to \infty$.
The approximate solution is expressed as 
${\bf a}_j = {\bf M}_j(\omega)~{\bf a}_0 = 0$ for sufficiently large $j$.
Requiring the existence of the solution with ${\bf a}_0 \neq 0$,
the relation 
\begin{align}
\det{\bf M}_j(\omega)= 0,
\label{eq:detmequal0}
\end{align}
should be satisfied for sufficiently large $j$, and 
the QNMs are approximately determined by this condition.
Let us discuss the quasinormal frequencies of our system~\eqref{eq: coupled master eq}.
For simplicity, we focus on $\ell=2$ and the fundamental modes. 
We expand the frequency $\omega$ around the lowest order as  
\begin{align}
\omega &= \frac{\Omega^{i}_{\rm Sch} }{2 M}+\epsilon^2\delta\omega_{i}~(i = {\rm s,~grav.}),
\label{eq:deltaomegai}
\end{align}
where $\Omega^{i}_{\rm Sch}~(i = {\rm s,~grav.})$ are 
the dimensionless Schwarzschild quasinormal frequencies normalized by the Schwarzschild radius
for the scalar field, the odd and even gravitational modes.\footnote{
Practically, we first calculate $\delta \bar{\omega}_i$ with $\omega = \Omega_{\rm Sch}^i/r_{\rm H} + \epsilon^2 \delta\bar{\omega}_i$,
and then we can obtain $\delta\omega_{i}$ in Eq.~\eqref{eq:deltaomegai} as
$\delta\omega_{i} = \delta\bar{\omega}_i - \delta R_{\rm H} \Omega_{\rm Sch}^i/(2M)$, where 
$r_{\rm H} = 2M(1 + \epsilon^2 \delta R_{\rm H})$.
}
For the fundamental modes, the values of $\Omega^{i}_{\rm Sch}$
are given by
$\Omega^{\rm s}_{\rm Sch} = 0.96728774 -0.19359239i$ and $\Omega^{\rm grav}_{\rm Sch} = 0.74734337 -0.17792463i$. 
Substituting these ansatzes into Eq.~\eqref{eq:detmequal0} in our system and expanding them with respect to $\epsilon$, we obtain the equations determining $\delta\omega_{i}$.

For the scalar field-led QNMs, {\it i.e.}, 
the case of $\omega = \Omega^{\rm s}_{\rm Sch}/(2 M)+\epsilon^2\delta\omega_{\rm s}$,
Eq.~\eqref{eq:detmequal0} becomes
\begin{align}
    F_0(\Omega^{\rm s}_{\rm Sch}) 
    +\epsilon^2 \{ F_{2,0}(\Omega^{\rm s}_{\rm Sch})
+
F_{2,1}(\Omega^{\rm s}_{\rm Sch})\delta\omega_{\rm s}
\}
    + {\cal O}(\epsilon^4) = 0,
    \label{eq:scalarqnmeq}
\end{align}
where $F_0, F_{2,0}$, and $F_{2,1}$ are functions of $\Omega^{\rm s}_{\rm Sch}$.
The lowest-order value $\Omega^{\rm s}_{\rm Sch}$ is determined by the equation $F_0=0$.
The second order ${\cal O}(\epsilon^2)$ expression of Eq.~\eqref{eq:scalarqnmeq} determines the value of $\delta\omega_{\rm s}$. 
We note that the term $F_{2,1}$ is the derivative of $F_0$ with respect to the frequency,
and 
the term $F_{2,0}$ is affected by the correction terms $V^{\rm s}_{\rm 2nd}$ and $V^{\rm s\pm}_{\rm 1st}$ in Eq.~\eqref{eq: potential_correction}.
The QNM correction $\delta\omega_{\rm s}$ is given by
\begin{align}
    (2M)\delta\omega_{\rm s} = b_1^2(45.225150 + 4.2484788 i) 
    + b_2^2(148.38628 + 20.29223 i).
\end{align}
Its limit of $b_2=0$ corresponds to the QNM in sGB gravity,\footnote{In Ref.~\cite{Blazquez-Salcedo:2016enn}, the authors investigate the GB term coupled with the exponential sector $\alpha e^{\phi}$, where $\alpha$ is a coupling constant.  In this case, an effective mass term appears in the scalar field equation. This mass term also modifies the QNMs of the scalar field at the order ${\cal O}(\alpha)$.} while that of $b_1=0$ corresponds to the QNM in dCS gravity~\cite{McManus:2019ulj}. 

\begin{figure}[tbp]
\begin{center}
\includegraphics[width=0.45\linewidth]{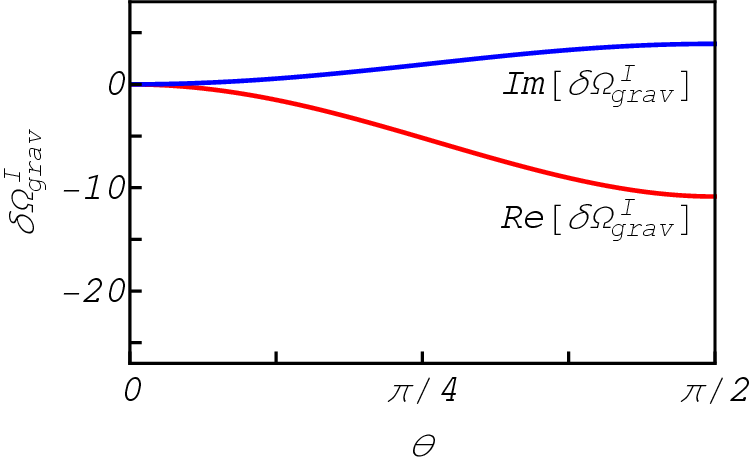}
\quad
\includegraphics[width=0.45\linewidth]{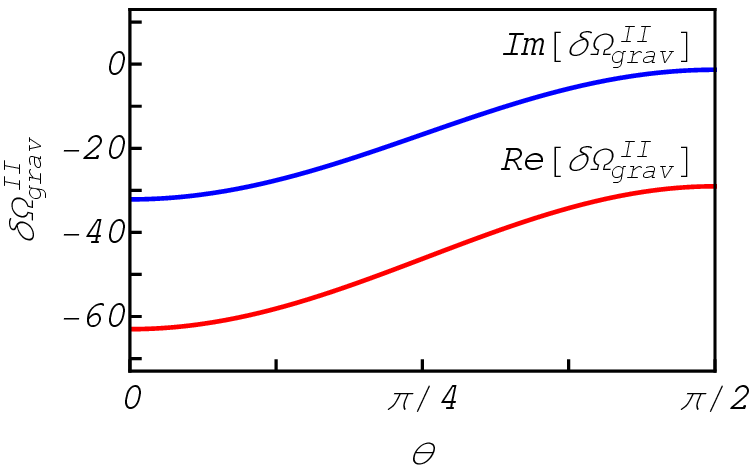}
\caption{
The dependence of 
$\delta\omega^{\rm I}_{\rm grav}$ and $\delta\omega^{\rm II}_{\rm grav}$ on $b_1$ and $b_2$.
Introducing $b_1 = B\sin{\theta}$ and $b_2 = B\cos{\theta}$,
we plot the values of 
the real and imaginary parts of $
\delta\Omega^{\rm I}_{\rm grav} :=
(2 M)\delta\omega^{\rm I}_{\rm grav}/B^2$ and 
$\delta\Omega^{\rm II}_{\rm grav} :=
(2 M)\delta\omega^{\rm II}_{\rm grav}/B^2$.
The dCS limit ($b_1 =0$) corresponds to $\theta = 0$ and
the sGB limit ($b_2 =0$) corresponds to $\theta = \pi/2$.
Since
$\delta\omega^{\rm I}_{\rm grav}$ and $\delta\omega^{\rm II}_{\rm grav}$
are functions of $b_1^2$ and $b_2^2$, 
we have only to consider the range $0 \le \theta \le \pi/2$.
}
\label{fig: delta omega I}
\end{center}
\end{figure}

The corrections to the lowest-order values of the odd and even gravitational modes are determined by the self-couplings of those modes $V^{\rm -}_{\rm 2nd}$ and $V^{\rm +}_{\rm 2nd}$, and the even-odd gravitation coupling $V^{\rm -+}_{\rm 2nd}$. 
Equation~\eqref{eq:detmequal0} is given by 
\begin{align}
&
    \tilde{F}_0(\Omega^{\rm grav}_{\rm Sch}) 
    +\epsilon^2 
    \{
        \tilde{F}_{2,0}(\Omega^{\rm grav}_{\rm Sch}) 
        +
    \tilde{F}_{2,1}(\Omega^{\rm grav}_{\rm Sch}) 
    \delta\omega_{\rm grav}
    \}
    \notag\\&\quad
    +\epsilon^4 \{
\tilde{F}_{4,0}(\Omega^{\rm grav}_{\rm Sch})
+
\tilde{F}_{4,1}(\Omega^{\rm grav}_{\rm Sch}) \delta\omega_{\rm grav}
+
\tilde{F}_{4,2}(\Omega^{\rm grav}_{\rm Sch}) \delta\omega_{\rm grav}^2
\}
    + {\cal O}(\epsilon^6)
    = 0.
    \label{eq:detmtensor}
\end{align}
The terms 
$\tilde{F}_0, \tilde{F}_{2,0}, \tilde{F}_{2,1}, \tilde{F}_{4,0},  \tilde{F}_{4,1},$ and $\tilde{F}_{4,2}$  are functions of $\Omega^{\rm grav}_{\rm Sch}$.
We solve the above equation order by order in $\epsilon$.
The lowest-order quasinormal frequency of the gravitational modes $\Omega^{\rm grav}_{\rm Sch}$ is
determined by $\tilde{F}_0 = 0$.
Since the spectra of even and odd gravitational modes are degenerate at the lowest order,
the terms $\tilde{F}_{2,0}$ and $F_{2,1}$, 
which are proportional to the first derivative of $\tilde{F}_0$, 
also vanish.
Thus, the QNM correction $\delta\omega_{\rm grav}$ is determined from 
Eq.~\eqref{eq:detmtensor} with ${\cal O}(\epsilon^4)$.
We note that 
the term $\tilde{F}_{4,2}$ is the second derivative of $2\tilde{F}_0$ with respect to the frequency,
and 
the terms $\tilde{F}_{4,0},  \tilde{F}_{4,1}$ are affected by the correction terms $V^{\rm \pm}_{\rm 2nd}, V^{\rm -+}_{\rm 2nd}$, and $V^{\rm s\pm}_{\rm 1st}$ in Eq.~\eqref{eq: potential_correction}.
The QNM correction $\delta\omega_{\rm grav}$ has two branches
\begin{align}
    &(2M)\delta\omega^{\rm I}_{\rm grav} = 
    (-5.3012524 +2.1911460 i)b_1^2
    -(31.506265+16.076674 i)b_2^2
\notag\\    
    &+ \sqrt{(75.825646 +47.759196i)b_1^4 +(532.00613 +469.85929i)b_1^2b_2^2 +(734.18529 +1013.0319i)b_2^4},
    \label{eq: omegaGrav correction I}
\\
    &(2M)\delta\omega^{\rm II}_{\rm grav} = 
    -(5.3012524 +2.1911460 i)b_1^2
    -(31.506265+16.076674 i)b_2^2
\notag\\    
    &-  \sqrt{(75.825646 +47.759196i)b_1^4 +(532.00613 +469.85929i)b_1^2b_2^2 +(734.18529 +1013.0319i)b_2^4}.
    \label{eq: omegaGrav correction II}
\end{align}
The existence of the two branches 
implies that the degeneracy of QNM spectra is broken at this order. 
The $b_1^2b_2^2$ term in the square root includes the effect of the new even-odd couplings $V^{\rm -+}_{\rm 2nd}$
and the scalar-gravity couplings $V^{\rm s\pm}_{\rm 1st}$.
Note that similar spectral features can be seen in the matrix toy models of Eq.~\eqref{eq:evlambda} in App.~\ref{sec: toy models}.
One can take the limits of the dCS and sGB gravity in Eqs.~\eqref{eq: omegaGrav correction I} and~\eqref{eq: omegaGrav correction II} by setting $b_1=0$
and $b_2 = 0$, respectively.\footnote{In the limit of the dCS gravity ($b_1=0$), the QNM frequencies are given by
\begin{align}
   {\rm odd~mode}:~&~ (2M) \delta\omega^{\rm II}_{\rm grav} 
 =  -b_2^2(63.012530 +32.153348 i),\label{eq: dCS odd}
 \\
  {\rm even~mode}:~&~  (2M) \delta\omega^{\rm I}_{\rm grav} 
 =0,\label{eq: dCS even}
\end{align}
while in the limit of the sGB gravity ($b_2=0$), those are given by
\begin{align}
    {\rm odd~mode}:~&~  (2M) \delta\omega^{\rm I}_{\rm grav} 
 = b_1^2(3.7937608 +0.43442445i),\label{eq: sGB odd}
\\
   {\rm even~mode}:~&~   (2M) \delta\omega^{\rm II}_{\rm grav} =-b_1^2(14.396266 +4.8167164i).\label{eq: sGB even}
\end{align}
We confirmed that 
these values
are consistent with the values for the weakly coupled regime in~\cite{Blazquez-Salcedo:2016enn,McManus:2019ulj}.
We note that 
the odd mode corresponds to 
$\delta\omega^{\rm II}_{\rm grav}$ for $b_1 =0$
but 
$\delta\omega^{\rm I}_{\rm grav}$ for $b_2 =0$,
and vice versa for the even mode.
This kind of branch exchange can be seen in the spectra of the matrix toy models in Eqs.~\eqref{eq:p0branch} and \eqref{eq:p1branch} of App.~\ref{sec: toy models}.
}
In Fig.~\ref{fig: delta omega I}, 
we plot the dependence of $\delta\omega^{\rm I}_{\rm grav}$ and 
$\delta\omega^{\rm II}_{\rm grav}$
on $b_1$ and $b_2$, where 
we introduce $b_1 = B\sin{\theta}, b_2 = B\cos{\theta}, \delta\Omega^{\rm I}_{\rm grav} :=
(2 M)\delta\omega^{\rm I}_{\rm grav}/B^2$ and 
$\delta\Omega^{\rm II}_{\rm grav} :=
(2 M)\delta\omega^{\rm II}_{\rm grav}/B^2$.
The limits of $\theta = 0$ and $\theta = \pi/2$ 
represent the cases of the dCS 
and sGB gravity, respectively.
From Fig.~\ref{fig: delta omega I}, for a fixed value of $B$, the decaying rates of both branches, {\it i.e.}, $-{\rm Im}[\delta\omega^{\rm I}_{\rm grav}]$ and $-{\rm Im}[\delta\omega^{\rm II}_{\rm grav}]$, decrease as $\theta$ increases.

\section{Rotating black hole solutions without $\mathbb{Z}_2$ symmetry}\label{sec: rotating BH}

In this section, we study the slowly rotating black holes up to the first order of the spin parameter.
We demonstrate that the parity-violating term in the dCS sector with the non-vanishing background value of the scalar field $\bar{\phi}$ leads to the $\mathbb{Z}_2$ violating rotating black holes.
For this purpose, we consider the $\ell=1$ time-independent perturbations around the static spherically symmetric black hole solutions (for example, see~\cite{Cardoso:2018ptl}). 
We note that $\ell \ge 2$ time-independent perturbations vanish at the first order of the spin parameter.
The ansatz of the $\ell=1$ time-independent perturbations is given by
\begin{align}
    \phi = \bar{\phi} +\delta\phi(r)\cos{\theta},~
    g_{\mu\nu} = \bar{g}_{\mu\nu} + \delta g_{\mu\nu},
    \label{eq: slow rot solution}
\end{align}
where the background solutions $\bar{\phi}$ and $\bar{g}_{\mu\nu}$ are given by Eqs.~\eqref{eq:gbar} and \eqref{eq:phibar} with Eqs.~\eqref{eq:agbar}--\eqref{eq: pi2nd}, and the form of $\delta g_{\mu\nu}$ is given by
\begin{align}
    \delta g_{\mu\nu}dx^{\mu}dx^{\nu}
    &= -2h_0\sin^2{\theta}dt d\varphi 
    +\cos{\theta}\bigg\{AH_0dt^2+2H_1dtdr
    +\frac{H_2}{B}dr^2\bigg\}.
    \label{eq: slow rot even}
\end{align}
We expand the perturbed quantities 
$\delta\phi,~h_0, H_1$ and $H_2$ as
\begin{align}
    \delta\phi(r) &= \delta\phi^{\rm 0th} + \epsilon\, \delta\phi^{\rm 1st} +\epsilon^2 \delta\phi^{\rm 2nd},
\\
    h_0(r) &= h^{\rm 0th}_0+\epsilon\, h^{\rm 1st}_0 +\epsilon^2 h^{\rm 2nd}_0,
    \label{eq: h0_l=1}
    \\
    H_1(r) &=  H^{\rm 0th}_1 + \epsilon\, H^{\rm 1st}_1 +\epsilon^2 H^{\rm 2nd}_1,
\\
    H_2(r) &= H^{\rm 0th}_2 + \epsilon\, H^{\rm 1st}_2 +\epsilon^2 H^{\rm 2nd}_2.
    \label{eq: slow rot expansion}
\end{align}
Solving the equations of motion \eqref{eq: scalarEoM} and \eqref{eq: gravityEoM} with the ansatz~\eqref{eq: slow rot solution}--\eqref{eq: slow rot expansion} perturbatively, we obtain slowly rotating black hole solutions. 
The lowest-order solution ${\cal O}(\epsilon^0)$ corresponds to the slowly rotating Kerr black hole up to the first order of the spin parameter $\chi$
\begin{align}
\delta\phi^{\rm 0th} &=0,\\
h^{\rm 0th}_0 &= \frac{r_{\rm g}}{r}\chi ,\\
H^{\rm 0th}_1 &=0,\\
H^{\rm 0th}_2 &=0.
\end{align}
The higher-order components up to ${\cal O}(\chi)$ are determined as
\begin{align}
    \delta\phi^{\rm 1st} &= \mpl b_2\chi \frac{  r_{\rm g} \left(10 r r_{\rm g}+9 r_{\rm g}^2+10 r^2\right)}{r^4},~
    \delta\phi^{\rm 2nd} =0,
\\
    h^{\rm 1st}_0 &= 0,
\\
    h^{\rm 2nd}_0 &= -b_1^2\chi \frac{2r_{\rm g}^3\left(140 r^3 r_{\rm g}+45 r^2 r_{\rm g}^2+36 r r_{\rm g}^3-50 r_{\rm g}^4+18 r^4\right)}{15
   r^7} 
\notag\\
    &\quad -b_2^2\chi 
   \frac{r_{\rm g}^4\left(240 r r_{\rm g}+189 r_{\rm g}^2+280 r^2\right)}{7 r^6},
\\
    H^{\rm 1st}_0 &= 0,~
    H^{\rm 2nd}_0 = \chi \frac{4b_1 b_2 r_{\rm g} \left(176 r^4 r_{\rm g}+243 r^3 r_{\rm g}^2+44 r^2 r_{\rm g}^3-57 r r_{\rm g}^4-144 r_{\rm g}^5+88
   r^5\right)}{7 r^7},
\\
    H^{\rm 1st}_1 &= 0,~H^{\rm 2nd}_1 = 0,
\\
    H^{\rm 1st}_2 &= 0,~
    H^{\rm 2nd}_2 = \chi \frac{4b_1 b_2 \left(388 r^4 r_{\rm g}^2 +477 r^3 r_{\rm g}^3-204 r^2 r_{\rm g}^4-507 r r_{\rm g}^5-768
   r_{\rm g}^6+264 r^5r_{\rm g}\right)}{7 r^7}.
\end{align}
The corresponding non-vanishing slowly rotating components of the perturbed metric are given by
\begin{align}
    \delta g_{tt} &= \epsilon^2\chi b_1 b_2\cos{\theta }\left(1-\frac{r_{\rm g}}{r}\right)\frac{4(176 r^4 r_{\rm g}^2+243 r^3 r_{\rm g}^3+44 r^2 r_{\rm g}^4-57 r r_{\rm g}^5-144 r_{\rm g}^6+88
   r^5r_{\rm g})}{7 r^7},\label{eq: dgtt}
\\
    \delta g_{rr} &=  \epsilon^2\chi \frac{b_1 b_2\cos{\theta }}{1-\frac{r_{\rm g}}{r}}\, \frac{4(388 r^4 r_{\rm g}^2+477 r^3 r_{\rm g}^3-204 r^2 r_{\rm g}^4-507 r r_{\rm g}^5-768
   r_{\rm g}^6+264 r^5r_{\rm g})}{7 r^7},\label{eq: dgrr}
\\
    \delta g_{t\phi} &= -\frac{r_{\rm g}}{r}\chi\sin^2{\theta}
    +b_1^2\epsilon^2\chi \sin^2{\theta}
   \frac{2r_{\rm g}^3 \left(140 r^3 r_{\rm g}+45 r^2 r_{\rm g}^2+36 r r_{\rm g}^3-50 r_{\rm g}^4+18 r^4\right)}{15 r^7}
\notag\\
    &\quad +b_2^2\epsilon^2\chi \sin^2{\theta}
    \frac{r_{\rm g}^4 \left(240 r r_{\rm g}+189 r_{\rm g}^2+280 r^2\right)}{ 7r^6}.
    \label{eq:slowrotgtphi}
\end{align}
Note again that
${\cal O}(\epsilon^0)$ term in Eq.~\eqref{eq:slowrotgtphi} represents the the slowly rotating Kerr solution. 
These slowly rotating solutions violate the $\mathbb{Z}_2$ symmetry $\theta \to \pi -\theta$ 
through the $\cos{\theta}$ terms in~\eqref{eq: dgtt} and \eqref{eq: dgrr}.
This kind of $\mathbb{Z}_2$ violation can be seen in the purely tensorial gravity theories with higher-curvature corrections~\cite{Cardoso:2018ptl,Cano:2019ore} and parity-violating modified gravity~\cite{Cunha:2018uzc, Tahara:2023pyg}.

\section{Mode mixing in tidal response}\label{sec: tidal}

Another effect of the new even-odd gravitational couplings is the tidal response from an external gravitational source.
The tidal response of a black hole is described by the ratio of growing and decaying modes to the static solution of the master equation, which is regular at the horizon.
This ratio is called the tidal Love number~\cite{Hinderer:2007mb,Binnington:2009bb,Damour:2009vw,Poisson:2014tb} and is useful to test gravity theories~\cite{Cardoso:2017cfl}. 
For simplicity, we focus on the $\ell = 2$ perturbations.
In the Schwarzschild limit $\epsilon = 0$, 
the $\ell=2$ static solutions of Eq.~\eqref{eq: coupled master eq} with the regularity at the horizon
are given by
\begin{align}
    \tilde{\Psi}^{\rm s}_{\rm Sch} &= C^{\rm s}
    \left(6\frac{r^3}{r_{\rm H}^3} -6\frac{r^2}{r_{\rm H}^2} +\frac{r}{r_{\rm H}}\right),\label{eq: 0th growing scalar}
\\
    \tilde{\Psi}^{\rm -}_{\rm Sch} &= C^{\rm -} \frac{r^3}{r_{\rm H}^3},\label{eq: 0th growing odd}
\\
    \tilde{\Psi}^{\rm +}_{\rm Sch} &= 
    \frac{C^{\rm +}}{4+3r_{\rm H}/r}
    \left(-4\frac{r^3}{r_{\rm H}^3} -6\frac{r^2}{r_{\rm H}^2} +3\right),\label{eq: 0th growing even}
\end{align}
where $C^{\rm s}$ and $C^\pm$ are integration constants, corresponding to the strength of the external tidal fields.
The absence of decaying modes in the far region in Eqs.~\eqref{eq: 0th growing scalar}--\eqref{eq: 0th growing even} implies vanishing Love numbers around Schwarzschild black holes~\cite{Binnington:2009bb}.
To determine the non-GR effect, 
we expand the static solutions of Eq.~\eqref{eq: coupled master eq} as
\begin{align}
    \tilde{\Psi}^{i} &= \tilde{\Psi}^{i}_{\rm Sch}
    +\epsilon \tilde{\Psi}^{i}_{\rm 1st}(r)
    +\epsilon^2 \tilde{\Psi}^{i}_{\rm 2nd}(r)~(i={\rm s},-,+).
    \label{eq:staticpsi}
\end{align}
Solving Eq.~\eqref{eq: coupled master eq} with the ansatz in Eq.~\eqref{eq:staticpsi},
we obtain the forms of $\tilde{\Psi}^{i}_{\rm 1st}$ and 
$\tilde{\Psi}^{i}_{\rm 2nd}$.
The explicit forms are shown in App.~\ref{sec: full response}. 
The new even-odd couplings yield 
\begin{align}
    \tilde{\Psi}^{-}_{\rm 2nd}|_{r\to \infty} &= -b_1 b_2 C^+ \left\{-\frac{1728}{7}-\frac{440}{7 }\frac{r_{\rm H}}{r}+\frac{6 (129024 \zeta (3)-154199)}{35}\frac{r_{\rm H}^2}{r^2}\right\}
\notag\\ 
    &\quad -b_1 b_2 C^+ \left(\frac{1536}{7}\frac{r_{\rm H}^2}{r^2} +\frac{1280}{7}\frac{r_{\rm H}^3}{r^3} \right) \ln \left(\frac{r}{r_{\rm H}}\right),\label{eq: -2nd}
\\
    \tilde{\Psi}^{+}_{\rm 2nd}|_{r\to \infty} &= 
    b_1 b_2 C^- \left\{-\frac{1728}{7} +\frac{208}{7}\frac{r_{\rm H}}{r} +\frac{6(129024 \zeta (3)-154439)}{35}\frac{r_{\rm H}^2}{r^2} \right\}
\notag\\ 
    &\quad +b_1 b_2 C^- \left(\frac{1536}{7}\frac{r_{\rm H}^2}{r^2}+\frac{512}{7}\frac{r_{\rm H}^3}{r^3}\right) \ln \left(\frac{r}{r_{\rm H}}\right),\label{eq: +2nd}
\end{align}
where $\zeta(3)$ is the Riemann zeta function $\zeta(n)$ with $n=3$.
This result shows that an even parity external tidal field $C^+$ induces an odd parity gravitational response in 
$\tilde{\Psi}^{-}_{\rm 2nd}$, and vice versa for the odd parity external tidal field $C^-$, as seen in~\cite{Cardoso:2018ptl}.
However, due to the presence of the logarithmic terms in Eqs~\eqref{eq: -2nd} and~\eqref{eq: +2nd}, it is not obvious how to define the tidal Love number.\footnote{
Note that the logarithmic terms appear in the worldline EFT of a black hole~\cite{Kol:2011vg,Hui:2020xxx,Ivanov:2022hlo,Charalambous:2022rre,Saketh:2023bul,Mandal:2023hqa} even in GR, modified gravity theories~\cite{Creci:2023cfx}, and gravitational EFTs~\cite{Charalambous:2022rre,DeLuca:2022tkm,Katagiri:2023umb,Bhattacharyya:2024aeq}.
In the world-line EFT, it is argued that these logarithmic terms can be cancelled out by taking into account ultraviolet physics in concrete models, and the tidal Love number becomes well-defined~\cite{Ivanov:2022hlo}.
}
Nevertheless, the tidal response through the even-odd gravitational couplings can be suggested.
The origin of the number $\zeta(3)$ in Eqs~\eqref{eq: -2nd} and~\eqref{eq: +2nd}
is the regularity condition at the horizon, 
so the asymptotic forms of $\tilde{\Psi}^{-}_{\rm 2nd}$ and $\tilde{\Psi}^{+}_{\rm 2nd}$
probably contain information about the horizon.
This suggests that 
$\tilde{\Psi}^{-}_{\rm 2nd}$ and $\tilde{\Psi}^{+}_{\rm 2nd}$ contain decaying modes, {\it i.e.}, the non-vanishing tidal responses,
because purely growing modes, which are determined only from the behaviour around the asymptotic region $r \to \infty$, are not expected to contain information about the horizon.

\section{Summary and discussion}\label{sec: summary}

In this paper, we investigate the effect of new even-odd gravitational couplings in the case where both the scalar Gauss-Bonnet (sGB) gravity sector and the dynamical Chern-Simons (dCS) gravity sector coexist. 
In the weak coupling regime, the effect of the new couplings appears as a correction to the general relativity (GR) solution.
Through the sGB gravity sector, the scalar field acquires a background value in static spherically symmetric black hole solutions, and as a result, the metric also deviates from the Schwarzschild solution while the dCS gravity sector does not affect it due to parity symmetry.
The master equations for gravitational perturbations around a static spherically symmetric solution are coupled among the scalar field and the odd and even gravitational modes of metric perturbations.
In particular, new even-odd gravitational couplings appear in the master equations.

To estimate the effect of the new couplings, the quasinormal frequencies around static spherically symmetric black holes, slowly rotating black holes, and tidal responses sourced by external gravitational objects were examined.
The tensor-led quasinormal frequencies are quantitatively affected by the new couplings because the explicit expressions contain the square root forms given in Eqs.~\eqref{eq: omegaGrav correction I} and~\eqref{eq: omegaGrav correction II}.
The appearance of the square root form is a specific feature of the eigenvalue problem when the lowest-order spectra are degenerate.
Thanks to the new even-odd couplings, the slowly rotating solution breaks the $\mathbb{Z}_2$ symmetry between the northern and southern hemispheres.
The static solution of the gravitational modes exhibits logarithmic terms due to the mediation of scalar fields. 
Because of the presence of these logarithmic terms, the tidal Love number cannot be well-defined. 
However, characteristic terms appear in the solutions after imposing regularity at the horizon, which suggests that the tidal response is induced by the new even-odd gravitational couplings.

Finally, we discuss the generality of even-odd gravitational couplings in scalar-tensor theories.
If we assume that the scalar field $\phi$ has a background value $\bar{\phi}$, 
the dCS gravity sector in the Lagrangian behaves as
\begin{align}
    \frac{b_2}{\Lambda}\phi\tilde{R}^{\mu\nu}{}_{\rho\sigma}R_{\nu\mu}{}^{\rho\sigma}
    \sim  \frac{b_2}{\Lambda}~\bar\phi~\partial^2h^{-}~\partial^2h^{+},\label{eq: dCS term}
\end{align}
where $h^\pm$ are the odd and even-type gravitational perturbations, respectively.
This demonstrates the existence of the even-odd gravitational couplings when $\bar{\phi} \neq 0$.
This structure is common in scalar-tensor theories with a single scalar field coupled to a parity-violating term other than the Pontryagin term.
While the even-odd gravitational couplings have already appeared in higher-curvature gravity theories~\cite{Cardoso:2018ptl,deRham:2020ejn,Cano:2021myl,Cano:2023jbk}, such couplings generically arise in scalar-tensor theories with parity-violating terms. 
The analysis of even-odd gravitational couplings in the general setting is left for future work~\cite{ours}.

\acknowledgements

We are grateful for fruitful discussions with Takuya Katagiri, and Kazufumi Takahashi.
This work was supported by 
the research fellowship at the National Institute of Technology, Oyama College (S.H.), JSPS KAKENHI Grant No. JP22K03626 (M.K.), and No. JP21H01080 (M.Y.), and 
IBS under the Project No. IBS-R018-D3 (M.Y.).
We acknowledge the hospitality at APCTP where part of this work was done during the focus research program ``Black Hole and Gravitational Waves: from modified theories of gravity to data analysis".

\appendix

\section{Explicit forms of master variables, deviations of sound speed and potentials}\label{sec: explicit forms of master eqs}

In Eqs.~\eqref{eq: coupled master eq}--\eqref{eq: potential_correction}, the coupled master equations of the scalar, odd, and even gravitational modes are derived. Due to the couplings, the master variables are mixed with other modes at higher orders in $\epsilon$. The relations of the master variables in the coupled master equation~\eqref{eq: coupled master eq} and those in Eqs.~\eqref{eq: master var scalar GR}--\eqref{eq: master var + GR} are given by 
\begin{align}
    &\tilde\Psi^{\rm s} =
        \frac{2 \sqrt{2} }{\sqrt{\lambda  (\lambda +2)} }\Psi^{\rm s}
\notag\\    
    &-b_1\epsilon
    \Bigg\{\frac{2 \sqrt{2}}{\sqrt{\lambda(\lambda +2) }(\lambda +3) r^4 (\lambda  r+3 r_{\rm g})} \Big(\lambda ^2 (\lambda +2) r^5+3 \lambda  (\lambda +2) r^4 r_{\rm g}+6 (\lambda +3) r^3 r_{\rm g}^2-6 (\lambda +3) r_{\rm g}^5\Big)
    \Psi^{+}
\notag\\
    &
     -\frac{4 \sqrt{2} r_{\rm g} (r^3-r_{\rm g}^3)}{\sqrt{\lambda  (\lambda +2)} r^3} \frac{\D\Psi^{+}}{\D r}\Bigg\}
\notag\\
    &
    -b_1^2\epsilon^2\Bigg\{\frac{16 \sqrt{2}}{\sqrt{\lambda(\lambda +2)}(\lambda +3)^2 r^6 (\lambda  r+3 r_{\rm g})}
    \Big(\lambda ^2 (\lambda +2) r^7+3 \lambda  (\lambda +2) r^6 r_{\rm g}+(\lambda ^2+9 \lambda +18) r^5 r_{\rm g}^2
\notag\\
    &
    +(2 \lambda ^3+11 \lambda ^2+18 \lambda +9) r^4 r_{\rm g}^3+(16 \lambda ^2+87 \lambda +117) r^3 r_{\rm g}^4-2 \lambda  (\lambda +3) r^2 r_{\rm g}^5+2 (\lambda +3)^2 (3 \lambda -1) r r_{\rm g}^6
\notag\\
    &    
    +4 (\lambda +3)^2 r_{\rm g}^7\Big)\Psi^{s}  
    -\frac{32 \sqrt{2} r_{\rm g}  (r^3-r_{\rm g}^3) (\lambda  r^3+3 r^2 r_{\rm g}+2 (\lambda +3) r_{\rm g}^3)}{\sqrt{\lambda } \sqrt{\lambda +2} (\lambda +3) r^5 (\lambda  r+3 r_{\rm g})}\frac{\D\Psi^{s}}{\D r}\Bigg\},
\\
    &\tilde\Psi^{-} =
        2\Psi^{-} +b_1^2\epsilon^2\frac{4 b_1^2 r_{\rm g}^2 (30 r^5+260 r^4 r_{\rm g}+135 r^3 r_{\rm g}^2+24 r^2 r_{\rm g}^3-582 r r_{\rm g}^4+280 r_{\rm g}^5)}{15 r^6 (r-r_{\rm g})}\Psi^{-}
\notag\\
    &
    -b_1 b_2\epsilon^2\Bigg\{\frac{ 4 r_{\rm g}^3(-60 \lambda  r^4-45 (\lambda +8) r^3 r_{\rm g}-3 (16 \lambda +45) r^2 r_{\rm g}^2+4 (25 \lambda -36) r r_{\rm g}^3+480 r_{\rm g}^4)}{15 r^6 (\lambda  r+3 r_{\rm g})}\Psi^{+}
\notag\\    
    & +\frac{16 r_{\rm g}^3 (r^3-r_{\rm g}^3)}{r^5}\frac{\D \Psi^{+}}{\D r}
    \Bigg\},
\end{align}
\begin{align}
    &\tilde\Psi^{+} = 
        \Psi^{+} +b_1\epsilon\frac{8 (\lambda  r^3+3 r^2 r_{\rm g}+2 (\lambda +3) r_{\rm g}^3)}{(\lambda +3) r^2 (\lambda  r+3 r_{\rm g})} \Psi^{\rm s}
\notag\\
    &
    +b_1^2\epsilon^2 \Bigg\{
    \frac{12}{10935 (\lambda +3)^2r^6 (\lambda  r+3 r_{\rm g})^2}\Big(
    -20 \lambda ^2 (2 \lambda ^7+9 \lambda ^6+27 \lambda ^4+243 \lambda ^3+1458 \lambda +4374) r^7 r_{\rm g}
\notag\\
    &    
    -7290 \lambda ^3 (\lambda +2) r^8 -90 \lambda  (\lambda  (\lambda  (\lambda  (2 \lambda ^4+9 \lambda ^3+27 \lambda +243)-162)-486)+1458) r^6 r_{\rm g}^2
\notag\\
    &    
    -60 \lambda  (\lambda  (\lambda  (2 \lambda ^4+9 \lambda ^3+27 \lambda +81)-1701)-4374) r^5 r_{\rm g}^3
\notag\\
    & 
    +45 (\lambda +3)^2 (\lambda  (\lambda  (\lambda  (2 \lambda -3)+243)+1188)+486) r^4 r_{\rm g}^4
\notag\\
    &
    -54 (\lambda +3) (\lambda  (\lambda  (\lambda  (2 \lambda -159)-1710)-6345)-5670) r^3 r_{\rm g}^5
\notag\\
    & +81 (\lambda +3) (\lambda  (\lambda  (92 \lambda +951)+3240)+5265) r^2 r_{\rm g}^6-972 (\lambda +3)^2 (41 \lambda -90) r r_{\rm g}^7-138510 (\lambda +3)^2 r_{\rm g}^8  \Big)\Psi^{+}
\notag\\    
    & +\frac{16 \lambda ^2 (\lambda +3) (2 \lambda -3) ((\lambda -3) \lambda +9)}{2187}\ln \bigg(\frac{\lambda  r+3 r_{\rm g}}{r}\bigg)\Psi^{+}
\notag\\
    & 
    +\frac{16 r_{\rm g} (r^3-r_{\rm g}^3) (\lambda  r^3+3 r^2 r_{\rm g}+2 (\lambda +3) r_{\rm g}^3)}{(\lambda +3) r^5 (\lambda  r+3 r_{\rm g})}\frac{\D\Psi^{+}}{\D r}
    \Bigg\}
\notag\\
    &+b_1 b_2\epsilon^2\Bigg\{\frac{16 r_{\rm g}^3(60 \lambda  r^4+15 (\lambda +24) r^3 r_{\rm g}+3 (4 \lambda +15) r^2 r_{\rm g}^2-4 (35 \lambda -9) r r_{\rm g}^3-600 r_{\rm g}^4)}{15 r^6 (\lambda  r+3 r_{\rm g})}\Psi^{-}
\notag\\
    &
   +\frac{64  r_{\rm g}^3 (r-r_{\rm g}) (r^2+r r_{\rm g}+r_{\rm g}^2)}{r^5}\frac{\D\Psi^{-}}{\D r}\Bigg\},
\end{align}
where we remind $\lambda = \ell^2+\ell-2$.
When the scalar field is allowed to have a background value, the propagation speed of each mode and the potential terms through the couplings are modified. The deviations of the sound speed of propagation modes in Eq.~\eqref{eq: sound speed} are given by
\begin{align}
    \Delta c_{\rm s}^2 &= -\frac{32 b_1^2 r_{\rm g}^2 (r+2 r_{\rm g})}{(\lambda +3) r^3},
\\
    \Delta c_{-}^2 &= -\frac{32 b_1^2 r_{\rm g}^3 (2 r^2+3 r r_{\rm g}+4 r_{\rm g}^2)}{r^5},
\\
     \Delta c_{+}^2 &= -\frac{32 b_1^2 r_{\rm g}^3 \{ 2\lambda^2 r^4+3 \lambda  (\lambda +6)r^3 r_{\rm g}+2 (\lambda +5) (2 \lambda +3) r^2 r_{\rm g}^2+(34\lambda +45) r r_{\rm g}^3+60 r_{\rm g}^4 \}}{r^5 (\lambda  r+3 r_{\rm g})^2}.
\end{align}
The potential term can be decomposed in $\epsilon$ order as seen in Eqs.~\eqref{eq: potential_GR} and~\eqref{eq: potential_correction}.
At the lowest order, they are given in Eqs.~\eqref{eq: KG potential}--\eqref{eq: Z potential}.
At first order, the scalar-odd and scalar-even couplings lead to the non-diagonal part of the potential in Eq.~\eqref{eq: Z potential}
\begin{align}
     V^{\rm s-}_{\rm1st} &= \frac{24 \sqrt{2\lambda  (\lambda +2)} b_2 r_{\rm g}^3}{r^5},
\\
    V^{\rm s+}_{\rm 1st} &= -\frac{4\sqrt{2\lambda  (\lambda +2)} b_1  r_{\rm g}^3 (3 \lambda ^2 r^2+13 \lambda  r r_{\rm g}+15 r_{\rm g}^2)}{r^5 (\lambda  r+3 r_{\rm g})^2}.
\end{align}
At the second order, 
the correction terms in Eq.~\eqref{eq: potential_correction} are given by
\begin{align}
    V^{\rm s}_{\rm 2nd} &= \frac{1152 b_2^2 (\lambda +2) r_{\rm g}^6}{r^8}
\notag\\
    &\quad +\frac{2 b_1^2 r_{\rm g}^2}{15 (\lambda +3) r^9 (\lambda  r+3 r_{\rm g})^2}\Big\{ -30 \lambda ^2 (\lambda ^2+3 \lambda +8) r^7+40 \lambda  (17 \lambda^3+28 \lambda ^2-21 \lambda -36) r^6 r_{\rm g}
\notag\\
    &\quad
-15 (7 \lambda ^4-321\lambda ^3-468 \lambda ^2+174 \lambda +144) r^5 r_{\rm g}^2-6 (16\lambda ^4+245 \lambda ^3-2239 \lambda ^2-1440 \lambda +450) r^4 r_{\rm g}^3
\notag\\
    &\quad +3 (690 \lambda ^4-458 \lambda ^3-8723 \lambda ^2+8013 \lambda -1530) r^3 r_{\rm g}^4+8 (2635 \lambda ^3+2661 \lambda ^2-16083\lambda +3267) r^2 r_{\rm g}^5
\notag\\
    &\quad +6 (12305 \lambda ^2+26823 \lambda -30276) r r_{\rm g}^6+86580 (\lambda +3) r_{\rm g}^7\Big\},
\\
    V^{\rm -}_{\rm 2nd} &=
    -\frac{2 b_1^2 r_{\rm g}^2}{15 r^9}\Big\{ 30 (\lambda -4) r^5+20 (38 \lambda -221) r^4 r_{\rm g}+15 (15 \lambda+146) r^3 r_{\rm g}^2+216 (\lambda -3) r^2 r_{\rm g}^3
\notag\\    
    &\quad -6 (205\lambda-5286) r r_{\rm g}^4-31140 r_{\rm g}^5
    \Big\},
\\
     V^{\rm +}_{\rm 2nd} &= -\frac{2 b_1^2 r_{\rm g}}{15 (\lambda +3) r^9 (\lambda  r+3 r_{\rm g})^4}\Big\{ 240 \lambda ^4 (\lambda +2) r^{10}+30 \lambda ^3 (\lambda ^3-\lambda^2+24 \lambda +48) r^9 r_{\rm g}
\notag\\
     &\quad +20 \lambda ^3 (74 \lambda ^3+415\lambda ^2+687 \lambda +72) r^8 r_{\rm g}^2+15 \lambda ^2 (15 \lambda^4+863 \lambda ^3+3646 \lambda ^2+5244 \lambda +612) r^7 r_{\rm g}^3
\notag\\
     &\quad +12 \lambda  (18 \lambda ^5+245 \lambda ^4+3943 \lambda ^3+9630\lambda ^2+6300 \lambda +1620) r^6 r_{\rm g}^4-6 (205 \lambda ^6+2417\lambda ^5+4188 \lambda ^4
\notag\\
     &\quad -17979 \lambda ^3-26190 \lambda ^2+24705 \lambda -2430) r^5 r_{\rm g}^5-12 (85 \lambda ^5+4987 \lambda^4+13893 \lambda ^3-11079 \lambda ^2
\notag\\
     &\quad -21735 \lambda +18225) r^4 r_{\rm g}^6+27 (560 \lambda ^4+3484 \lambda ^3+4717 \lambda ^2+975 \lambda+9180) r^3 r_{\rm g}^7-72 (1240 \lambda ^3
\notag\\
     &\quad -3963 \lambda ^2-22734\lambda +945) r^2 r_{\rm g}^8-162 (3225 \lambda ^2+5671 \lambda -12012) r r_{\rm g}^9-604260 (\lambda +3) r_{\rm g}^{10}
     \big\},
\end{align}
\begin{align}
    V^{\rm -+}_{\rm 2nd} &= \frac{16 b_1 b_2 r_{\rm g}^3}{5 (\lambda +3) r^9 (\lambda  r+3 r_{\rm g})^2} \Big\{ 30 \lambda ^2 (\lambda ^2+14 \lambda +36) r^6+10 \lambda  (-17 \lambda ^2+87 \lambda +468) r^5 r_{\rm g}
\notag\\
    &\quad -30 (40 \lambda ^2+69 \lambda -180) r^4 r_{\rm g}^2-21 (90 \lambda ^3+271 \lambda ^2+78 \lambda +225) r^3 r_{\rm g}^3
\notag\\
    &\quad +2 (1000 \lambda ^3-1838 \lambda ^2-14541 \lambda -81) r^2 r_{\rm g}^4+294 (35 \lambda ^2+62 \lambda -129) r r_{\rm g}^5+13440 (\lambda +3) r_{\rm g}^6 \Big\}.
\end{align}
The master variables are related to the original perturbed scalar field in Eq.~\eqref{eq: BGscalar+pert} and the metric perturbations in Eq.~\eqref{eq: BGmetric+pert} as 
\begin{align}
    &\frac{r\Pi}{\mpl } =
    \frac{\sqrt{\lambda  (\lambda +2)}}{2 \sqrt{2}}\tilde\Psi^s
\notag\\
    & 
    +b_1 \epsilon\Bigg\{\frac{\lambda ^2 (\lambda +2) r^5+3 \lambda  (\lambda +2) r^4 r_{\rm g}+6 (\lambda +3) r^3 r_{\rm g}^2-6 (\lambda +3) r_{\rm g}^5}{(\lambda +3) r^4 (\lambda  r+3 r_{\rm g})}\tilde\Psi^+
    -2 r_{\rm g} \left(1-\frac{r_{\rm g}^3}{r^3}\right) \frac{\D\tilde\Psi^+}{\D r}\Bigg\}
\notag\\
    & +b_1^2 \epsilon^2\frac{2 \sqrt{2} \sqrt{\lambda  (\lambda +2)} r_{\rm g}^2}{(\lambda +3) r^6 (\lambda  r+3 r_{\rm g})}\Big\{\lambda  r^5+(\lambda +3) r^4 r_{\rm g}+(4 \lambda +3) r^3 r_{\rm g}^2-2 (\lambda -3) r^2 r_{\rm g}^3
\notag\\
    & +2 (3 \lambda ^2+8 \lambda -3) r r_{\rm g}^4+16 (\lambda +3) r_{\rm g}^5\Big\}\tilde\Psi^+,
\end{align}
\begin{align}
    &h_0 = \frac{r-r_{\rm g}}{2r}\left(\tilde{\Psi}^{-} +r\frac{\D\tilde\Psi^-}{\D r}\right)
\notag\\
    & +b_1^2\epsilon^2\Bigg\{\frac{r_{\rm g}^2 (30 r^5+980 r^4 r_{\rm g}-465 r^3 r_{\rm
g}^2+84 r^2 r_{\rm g}^3-4122 r r_{\rm g}^4+3640 r_{\rm g}^5)}{15 r^7}\tilde\Psi^-
\notag\\
    & +\frac{r_{\rm g}^2 (30 r^5+260 r^4 r_{\rm g}+15 r^3 r_{\rm g}^2+24 r^2 r_{\rm g}^3-582 r r_{\rm g}^4+400 r_{\rm g}^5)}{15 r^6}\frac{\D\tilde\Psi^-}{\D r}\Bigg\}
\notag\\
    & +b_1b_2\epsilon^2\Bigg\{\frac{4 r_{\rm g}^3}{15 r^7} \Bigg[\frac{r-r_{\rm g}}{(\lambda  r+3 r_{\rm g})^2}\Big(360 \lambda ^2 r^5+45 \lambda  (3 \lambda +44) r^4 r_{\rm g}+6 (22 \lambda ^2+75 \lambda +360) r^3 r_{\rm g}^2
\notag\\    
    & +(-1220 \lambda ^2+432 \lambda +135) r^2 r_{\rm g}^3-12 (580 \lambda -9) r r_{\rm g}^4-9360 r_{\rm g}^5\Big)
\notag\\
    & -60 r^4 (r^2+rr_{\rm g}+r_{\rm g}^2) \left(\omega ^2-\frac{(r-r_{\rm g}) (\lambda ^2 \
(\lambda +2) r^3+3 \lambda ^2 r^2 r_{\rm g}+9 \lambda  r r_{\rm \
g}^2+9 r_{\rm g}^3)}{r^4 (\lambda  r+3 r_{\rm g})^2}\right)\Bigg]\tilde\Psi^+
\notag\\
    &  +\frac{4 r_{\rm g}^3 (r-r_{\rm g})}{15 r^6 (\lambda  r+3 r_{\rm g})}\Big[120 \lambda  r^4+15 (\lambda +12) r^3 r_{\rm g}+3 (4 \lambda +15) r^2 r_{\rm g}^2
    -4 (50 \lambda -9) r r_{\rm g}^3-420 r_{\rm g}^4\Big]\frac{\D\tilde\Psi^+}{\D r}\Bigg\},\label{eq: h0}
\end{align}
\begin{align}
    &h_1 = -\frac{i r^2 \omega }{2 (r-r_{\rm g})}\tilde{\Psi}^{-}
\notag\\
    & +b_1^2\epsilon^2\frac{i r_{\rm g}^2 \omega }{15 r^4 (r-r_{\rm g})^2}(30 r^5+260 r^4 r_{\rm g}+135 r^3 r_{\rm g}^2+24 r^2 r_{\rm g}^3-582 r r_{\rm g}^4+280 r_{\rm g}^5)\tilde\Psi^-
\notag\\
     & +b_1b_2\epsilon^2\Bigg\{\frac{4 i r_{\rm g}^3 \omega }{15 r^4 (r-r_{\rm g}) (\lambda  r+3 r_{\rm g})}\Big[60 \lambda  r^4+45 (\lambda +8) r^3 r_{\rm g}+3 (16 \lambda +45) r^2 r_{\rm g}^2-4 (25 \lambda -36) r r_{\rm g}^3 -480 r_{\rm g}^4\Big]\tilde\Psi^+
\notag\\
    & 
    -\frac{16 i r_{\rm g}^3 \omega  (r^2+r r_{\rm g}+r_{\rm g}^2)}{r^3}\frac{\D \tilde\Psi^+}{\D r}\Bigg\},\label{eq: h1}
\end{align}
\begin{align}
    &H_0 = 
    \Bigg\{\frac{\lambda ^2 (\lambda +2) r^3+3 \lambda ^2 r^2 r_{\rm g}+9 \lambda  r r_{\rm g}^2+9 r_{\rm g}^3}{2 r^2 (\lambda  r+3 r_{\rm g})^2} -\frac{r^2 \omega ^2}{r-r_{\rm g}}\Bigg\}\tilde\Psi^-
    -\frac{3 \lambda  r r_{\rm g} -2 \lambda  r^2+3 r_{\rm g}^2}{2 
\lambda  r^2+6 r r_{\rm g}}\frac{\D \tilde\Psi^-}{\D r}
\notag\\
    & +b_1\epsilon\Bigg\{\frac{2 \sqrt{2}\sqrt{\lambda  (\lambda +2)} r^2 \omega ^2}{(\lambda +3) (r-r_{\rm g})}\tilde\Psi^{\rm s} 
    -\frac{\sqrt{2} \sqrt{\lambda  (\lambda +2)}}{(\lambda +3) r^4(\lambda  r+3 r_{\rm g})^2}\Bigg[
    \Big(\lambda ^2 (\lambda +2) r^5+\lambda  (7 \lambda +12) r^4 r_{\rm g}
\notag\\
    & +3 (5 \lambda +6) r^3 r_{\rm g}^2+(4 \lambda ^3+12 \lambda ^2+9) r^2 r_{\rm g}^3+14 \lambda  (\lambda +3) r r_{\rm g}^4+12 (\lambda +3) r_{\rm g}^5\Big)\tilde\Psi^{\rm s}
\notag\\
    & +r \Big(2 \lambda ^2 r^5-(\lambda -12) \lambda  r^4 r_{\rm g}-6 (\lambda -3) r^3 r_{\rm g}^2-9 r^2 r_{\rm g}^3-2 \lambda  (\lambda +3) r r_{\rm g}^4
    -6 (\lambda +3) r_{\rm g}^5\Big)
\frac{\D\tilde\Psi^{\rm s}}{\D r}\Bigg]\Bigg\}
\notag\\ 
    & +b_1^2\epsilon^2\Bigg\{-\frac{r_{\rm g}}{5 (\lambda +3) r^8 (\lambda  r+3 r_{\rm g})^4}\Big[80 \lambda ^4 (\lambda +2) r^{10}-40 \lambda ^3 (2 \lambda ^2-3\lambda -12) r^9 r_{\rm g}
\notag\\ 
    & +10 \lambda ^3 (16 \lambda ^3+99 \lambda^2+141 \lambda -60) r^8 r_{\rm g}^2
    -20 \lambda ^2 (\lambda ^4-33\lambda ^3-178 \lambda ^2-306 \lambda +36) r^7 r_{\rm g}^3
\notag\\ 
    & -5 \lambda^2 (4 \lambda ^4+71 \lambda ^3-75 \lambda ^2+24 \lambda +756) r^6 r_{\rm g}^4
    -4 \lambda  (25 \lambda ^5+423 \lambda ^4+1316 \lambda^3-1089 \lambda ^2-3375 \lambda +4050) r^5 r_{\rm g}^5
\notag\\ 
    & +2 (235 \lambda^5-1295 \lambda ^4-5817 \lambda ^3+11889 \lambda ^2+32535 \lambda-1215) r^4 r_{\rm g}^6
\notag\\     
    & +24 (110 \lambda ^4+754 \lambda ^3+1602 \lambda^2+1845 \lambda +2565) r^3 r_{\rm g}^7
    +3(-3580 \lambda ^3+144\lambda ^2+36027 \lambda +10125) r^2 r_{\rm g}^8
\notag\\     
    & -108 (415 \lambda^2+1127 \lambda -354) r r_{\rm g}^9-35370 (\lambda +3) r_{\rm g}^{10}\Big]\tilde\Psi^+
\notag\\
    & +\frac{2r_{\rm g}^3\omega^2}{15 r^4 (r-r_{\rm g})^2 (\lambda  r+3 r_{\rm g})^2}\Big[10 \lambda ^2 r^6+10 \lambda  (19 \lambda +66) r^5 r_{\rm g}+3 (11 \lambda ^2+180 \lambda +270) r^4 r_{\rm g}^2
\notag\\
    & +6 (4 \lambda^2+33 \lambda +165) r^3 r_{\rm g}^3+(-110 \lambda ^2-816 \lambda +297) r^2 r_{\rm g}^4+12 (25 \lambda -132) r r_{\rm g}^5 +810 r_{\rm g}^6\Big]\tilde\Psi^+\Bigg\}
\notag\\
    & +b_1b_2\epsilon^2\Bigg\{-\frac{4 r_{\rm g}^3}{15 r^8 (\lambda  r+3 r_{\rm g})^2}\Big[
    60 (\lambda -16) \lambda ^2 r^6+15 \lambda  (\lambda ^2+118 \lambda -144) r^5r_{\rm g}
    +3 \lambda  (4 \lambda ^2+63 \lambda +1740) r^4 r_{\rm g}^2
\notag\\
    & +(-140 \lambda ^3+3116 \lambda ^2+855 \lambda +3240) r^3r_{\rm g}^3
    +3 (-1700 \lambda ^2+2796 \lambda +405) r^2 r_{\rm g}^4
\notag\\
    & -36(460 \lambda -33) r r_{\rm g}^5-10980 r_{\rm g}^6\Big]\tilde\Psi^-
    -\frac{8 r_{\rm g}^3 \omega ^2 (-120 r^3+15 r^2 r_{\rm g}+18 r r_{\rm g}^2+230 r_{\rm g}^3)}{15 r^4 (r-r_{\rm g})}\tilde\Psi^-
\notag\\
    & +\Bigg[-\frac{4 r_{\rm g}^3}{15 r^7 (\lambda  r+3 r_{\rm g})}(60 (\lambda -4) \lambda  r^5+450 \lambda  r^4 r_{\rm g}+3 (13 \lambda +60) r^3 r_{\rm g}^2
    +(-60 \lambda ^2+464 \lambda +135) r^2 r_{\rm g}^3
\notag\\
    & +12 (12-55 \lambda ) r r_{\rm g}^4-300 r_{\rm g}^5)
    +\frac{32 r_{\rm g}^3 \omega ^2 (r^2+r r_{\rm g}+r_{\rm g}^2)}{r^3}\Bigg]\frac{\D\tilde\Psi^-}{\D r}\Bigg\},
\end{align}
\begin{align}
    &H_1 = \frac{i \omega  (-2 \lambda  r^2+3 \lambda  r r_{\rm g}+3 r_{\rm g}^2)}{2 (r-r_{\rm g}) (\lambda  r+3 r_{\rm g})}\tilde\Psi^+
    -i r \omega\frac{\D\tilde\Psi^+}{\D r}
\notag\\
    & +b_1\epsilon\frac{i \sqrt{2\lambda  (\lambda +2)} \omega }{(\lambda +3) r^2 (r-r_{\rm g}) (\lambda  r+3 r_{\rm g})}\Bigg\{\Big[2 \lambda  r^4-(\lambda -6) r^3 r_{\rm g}-3 r^2 r_{\rm g}^2-2 (\lambda +3) r_{\rm g}^4\Big]\tilde\Psi^+ 
\notag\\
    & +2 r^3 (r-r_{\rm g}) (\lambda  r+3 r_{\rm g})\frac{\D\tilde\Psi^+}{\D r}\Bigg\}
\notag\\
    & +b_1^2\epsilon^2\Bigg\{\frac{i r_{\rm g} \omega }{15 (\lambda +3) r^6 (r-r_{\rm g})^2 (\lambda  r+3 r_{\rm g})^3}[120 \lambda ^3 (\lambda +2) r^{10}-120 \lambda ^2 (\lambda ^2-2\lambda -6) r^9 r_{\rm g}
\notag\\
    & +30 \lambda ^2 (47 \lambda ^2+131 \lambda \
+6) r^8 r_{\rm g}^2-10 \lambda  (290 \lambda ^3+297 \lambda ^2-1701 \
\lambda -162) r^7 r_{\rm g}^3
\notag\\
    & +15 \lambda  (67 \lambda ^3-949 \lambda \
^2-3552 \lambda -522) r^6 r_{\rm g}^4
    -3 (964 \lambda ^4+681 \lambda^3+1377 \lambda ^2+27540 \lambda +13770) r^5 r_{\rm g}^5
\notag\\
    & +(5990\lambda ^4+9696 \lambda ^3-14643 \lambda ^2+22437 \lambda -21060) r^4r_{\rm g}^6
\notag\\
    & +3 (-920 \lambda ^4+5466 \lambda ^3+32964 \lambda ^2+26343\lambda +4455) r^3 r_{\rm g}^7
    -6 (2170 \lambda ^3+8211 \lambda
^2-5643 \lambda -32238) r^2 r_{\rm g}^8
\notag\\
    & -90 (38 \lambda ^2+981 \lambda+2601) r r_{\rm g}^9+25920 (\lambda +3) r_{\rm g}^{10}]\tilde\Psi^+
    -\frac{32 i r_{\rm g}^4 \omega ^3 (r^2+r r_{\rm g}+r_{\rm g}^2)}{r^2(r-r_{\rm g}) (\lambda  r+3 r_{\rm g})}\tilde\Psi^+
\notag\\
    & +\frac{8 i r_{\rm g}^4 \omega}{r^5 (\lambda  r+3 r_{\rm g})^2}\Big[(\lambda -2) \lambda  r^4+(\lambda ^2+2 \lambda -12) r^3 r_{\rm g}+(\lambda ^2+2 \lambda -3) r^2 r_{\rm g}^2
    +(10 \lambda -3) r r_{\rm g}^3+27 r_{\rm g}^4\Big]\frac{\D\tilde\Psi^+}{\D r}\Bigg\}
\notag\\
    & +b_1b_2\epsilon^2\Bigg\{-\frac{4 i r_{\rm g}^3 \omega }{15 r^6 (r-r_{\rm g}) (\lambda  r+3r_{\rm g})}\Big[-120 (\lambda -6) \lambda  r^5-30 (23 \lambda -48) r^4 r_{\rm g}+3 (7\lambda -120) r^3 r_{\rm g}^2
\notag\\
    & +(120 \lambda ^2-2684 \lambda +45) r^2r_{\rm g}^3+12 (215 \lambda -567) r r_{\rm g}^4+5520 r_{\rm g}^5\Big]\tilde\Psi^-
    -\frac{32 i r_{\rm g}^3 \omega ^3 (r^2+r r_{\rm g}+r_{\rm g}^2)}{r^2 (r-r_{\rm g})}\tilde\Psi^-
\notag\\
    & -\frac{8 i r_{\rm g}^4 \omega  (75 r^2+78 r r_{\rm g}-10 r_{\rm g}^2)}{15 r^5}\frac{\D\tilde\Psi^-}{\D r}\Bigg\},
\end{align}
\begin{align}
    &H_2 = \Bigg\{\frac{\lambda ^2 (\lambda +2) r^3+3 \lambda ^2 r^2 r_{\rm g}+9 \lambda  r r_{\rm g}^2+9 r_{\rm g}^3}{2 r^2 (\lambda  r+3 r_{\rm g})^2}-\frac{r^2 \omega ^2}{r-r_{\rm g}}\Bigg\}\tilde\Psi^+
    -\frac{-2 \lambda  r^2+3 \lambda  r r_{\rm g}+3 r_{\rm g}^2}{2 \lambda  r^2+6 r r_{\rm g}}\frac{\D \tilde\Psi^+}{\D r}
\notag\\
    & +b_1\epsilon\Bigg\{\frac{2\sqrt{2\lambda  (\lambda +2)} r^2 \omega ^2}{(\lambda +3) (r-r_{\rm g})}\tilde\Psi^{\rm s}
    -\frac{\sqrt{2\lambda (\lambda +2)}}{(\lambda +3) r^4(\lambda  r+3 r_{\rm g})^2}\Bigg[(\lambda ^2 (\lambda +2) r^5+\lambda  (7 \lambda +12) r^4 r_{\rm g}
\notag\\
    & +3(5 \lambda +6) r^3 r_{\rm g}^2+(8 \lambda ^3+24 \lambda ^2+9) r^2 r_{\rm g}^3+38 \lambda  (\lambda +3) r r_{\rm g}^4+48 (\lambda +3) r_{\rm g}^5)\tilde\Psi^{\rm s}
 \notag\\
    & +r \Big(2 \lambda ^2 r^5-(\lambda -12) \lambda  r^4 r_{\rm g}-6 (\lambda -3) r^3 r_{\rm g}^2-9 r^2 r_{\rm g}^3-2 \lambda  (\lambda +3) r r_{\rm g}^4-6 (\lambda +3) r_{\rm g}^5\Big)\frac{\D\tilde\Psi^{\rm s}}{\D r}\Bigg]\Bigg\}
 \notag\\
    & +b_1^2\epsilon^2\Bigg\{-\frac{r_{\rm g}}{5 (\lambda +3) r^8 (\lambda  r+3 r_{\rm g})^4} \Big[80 \lambda ^4 (\lambda +2) r^{10}-40 \lambda ^3 (2 \lambda ^2-3\lambda -12) r^9 r_{\rm g}
\notag\\
    & +10 \lambda^3 (40 \lambda ^3+195 \lambda^2+237 \lambda -60) r^8 r_{\rm g}^2+20 \lambda ^2 (3 \lambda ^4+173\lambda ^3+610 \lambda ^2+594 \lambda -36) r^7 r_{\rm g}^3
\notag\\
    & +15 \lambda^2 (4 \lambda ^4+51 \lambda ^3+969 \lambda ^2+2200 \lambda +324) r^6r_{\rm g}^4
\notag\\
    & -4\lambda  (105 \lambda ^5+643 \lambda ^4+296 \lambda^3-11169 \lambda ^2-22815 \lambda +4050) r^5 r_{\rm g}^5
\notag\\
    & -2 (1205\lambda ^5+6095 \lambda ^4+2937 \lambda ^3-45369 \lambda ^2-90855\lambda +1215) r^4 r_{\rm g}^6
\notag\\
    & -24 (390 \lambda ^4+686 \lambda ^3-2232\lambda ^2-4275 \lambda -5805) r^3 r_{\rm g}^7
    +3 (-14620 \lambda^3-29376 \lambda ^2+48987 \lambda +16605) r^2 r_{\rm g}^8
\notag\\
    & -108 (935\lambda ^2+2627 \lambda -534) r r_{\rm g}^9 -74250 (\lambda +3) r_{\rm g}^{10}\Big]\tilde\Psi^+
\notag\\
    & +\frac{2 r_{\rm g}^3 \omega ^2}{15 r^4 (r-r_{\rm g})^2 (\lambda  r+3 r_{\rm g})^2}\Big[490 \lambda ^2 r^6+10 (354-5 \lambda ) \lambda  r^5 r_{\rm g}+3 (11\lambda ^2-300 \lambda +1710) r^4 r_{\rm g}^2
\notag\\
    & -6 (196 \lambda ^2-33\lambda +195) r^3 r_{\rm g}^3+(850 \lambda ^2-8016 \lambda +297) r^2r_{\rm g}^4+12 (505 \lambda -1032) r r_{\rm g}^5+9450 r_{\rm g}^6\Big]\tilde\Psi^+
\notag\\
    & -\frac{r_{\rm g}}{5 (\lambda +3) r^7 (\lambda  r+3 r_{\rm g})^3}\Big[40 \lambda ^3 (\lambda +2) r^9-40 \lambda ^2 (\lambda ^2-\lambda -6) r^8 r_{\rm g} +10 \lambda ^2 (47 \lambda ^2+123 \lambda -42) r^7 r_{\rm g}^2
\notag\\
    & -10 \lambda  (88 \lambda ^3+29 \lambda ^2-675 \lambda +54) r^6 r_{\rm g}^3-5 \lambda  (19 \lambda ^3+1303 \lambda ^2+3180 \lambda -1530) r^5 r_{\rm g}^4
\notag\\
    & -(1052 \lambda ^4+3935 \lambda ^3+16587 \lambda ^2+42120 \lambda -810) r^4 r_{\rm g}^5
    +(1470 \lambda ^4-1310 \lambda ^3-19287 \lambda ^2-17541 \lambda -33480) r^3 r_{\rm g}^6
\notag\\
    & +(9950 \lambda ^3+21762 \lambda ^2-26199 \lambda -5805) r^2 r_{\rm g}^7+6 (3565 \lambda ^2+10377 \lambda -954) r r_{\rm g}^8
    +15030 (\lambda +3) r_{\rm g}^9\Big]\frac{\D\tilde\Psi^+}{\D r}
\notag\\
    & +\frac{32 r_{\rm g}^4 \omega ^2 (r^2+r r_{\rm g}+r_{\rm g}^2)}{r^3 (\lambda  r+3 r_{\rm g})}\frac{\D\tilde\Psi^+}{\D r}\Bigg\}
\notag\\
    & +b_1b_2\epsilon^2\Bigg\{\frac{4 r_{\rm g}^3}{15 r^8 (\lambda  r+3 r_{\rm g})^2}\Big[180 \lambda ^2 (\lambda +8) r^6+15 \lambda  (7 \lambda ^2-54 \lambda +336) r^5 r_{\rm g} +36 (580 \lambda -363) r r_{\rm g}^5
\notag\\
    & +3 (36 \lambda ^3+137 \lambda ^2-1980 \lambda +1440) r^4 r_{\rm g}^2-(340 \lambda ^3+3716 \lambda ^2+495 \lambda +7560) r^3 r_{\rm g}^3
\notag\\
    & +3 (1220 \lambda ^2-5076 \lambda -765) r^2 r_{\rm g}^4 +23940 r_{\rm g}^6\Big]\tilde\Psi^-
    -\frac{8 r_{\rm g}^3 \omega ^2 (120 r^3+135 r^2 r_{\rm g}+138 r r_{\rm g}^2-250 r_{\rm g}^3)}{15 r^4 (r-r_{\rm g})}\tilde\Psi^-
\notag\\
    & -\frac{4 r_{\rm g}^3}{15 r^7 (\lambda  r+3 r_{\rm g})}\Big[60 (\lambda -12) \lambda  r^5+30 (31 \lambda -48) r^4 r_{\rm g}+3 (53\lambda +540) r^3 r_{\rm g}^2 +12 (342-175 \lambda ) r r_{\rm g}^4
\notag\\
    & +(-60 \lambda ^2+1784 \lambda +495) r^2r_{\rm g}^3 -4620 r_{\rm g}^5\Big]\frac{\D\tilde\Psi^-}{\D r}
    +\frac{32 r_{\rm g}^3 \omega ^2 (r^2+r r_{\rm g}+r_{\rm g}^2)}{r^3}\frac{\D\tilde\Psi^-}{\D r}\Bigg\}.
\end{align}

\section{Perturbative eigenvalues of symmetric matrix}\label{sec: toy models}

In this section, we study the eigenvalues of symmetric matrices that are diagonal at the lowest order but have non-diagonal components perturbatively.
Since the characteristics of the perturbative spectrum of eigenvalues of symmetric matrices are similar to those of the spectrum of quasinormal frequencies of black holes, this study is useful for understanding the results of Sec.~\ref{sec: QNM}.

\subsection{$2 \times 2$ case: non-degenerate}

We consider a toy model in which there are two components with a mixing term characterized by a small parameter $\delta$
\begin{align}
    {\bf M} = \begin{pmatrix}
    1 & 0  \\
    0 & 2  \\
    \end{pmatrix}
    +
    \begin{pmatrix}
    0 & \delta  \\
    \delta & 0  \\
    \end{pmatrix},
    \label{eq:matrixnd}
\end{align}
and the spectra are not degenerate at the lowest order.
For simplicity, let the lowest order eigenvalues be $1$ and $2$ because this choice does not lose the essence of the argument.
This model mimics the spectral problems of a coupled system of a single scalar mode and a single gravitational mode.
The eigenvalues $\lambda$ are determined by the characteristic equation
\begin{align}
{\rm Det}[{\bf M} - \lambda {\bf I}] = 0,
\label{eq:matrixeigenvalueeq}
\end{align}
where ${\bf I}$ is the unit matrix,
and this equation for Eq.~\eqref{eq:matrixnd} becomes
\begin{align}
    (1-\lambda)(2-\lambda)-\delta^2 =0.
    \label{eq:lambdaeqfornd}
\end{align}
Substituting $\lambda=1 +\Delta\lambda$ into Eq.~\eqref{eq:lambdaeqfornd}, we obtain
$-\Delta \lambda - \delta^2 = 0$ at ${\cal O}(\Delta \lambda)$.
Thus, the corresponding eigenvalue 
is $\lambda = 1 -\delta^2 +{\cal O}(\delta^3)$.
On the other hand, substituting $\lambda = 2 +\Delta\lambda$ 
into Eq.~\eqref{eq:lambdaeqfornd}, 
we obtain
$\Delta \lambda - \delta^2 = 0$ at ${\cal O}(\Delta \lambda)$,
and the corresponding eigenvalue becomes $\lambda = 2 +\delta^2 +{\cal O}(\delta^3)$.
For symmetric matrices which are not degenerate at the lowest order,
we can see that the correction of the eigenvalue appears at ${\cal O}(\delta^2)$.

\subsection{$2 \times 2$ case: degenerate}

We consider a matrix
whose spectra are degenerate at the lowest order
\begin{align}
    {\bf M} = \begin{pmatrix}
    2 & 0  \\
    0 & 2  \\
    \end{pmatrix}
    +
    \begin{pmatrix}
    0 & \delta  \\
    \delta & 0  \\
    \end{pmatrix},
    \label{eq:matrixdeg}
\end{align}
where we assume the lowest order eigenvalues to be $2$ for simplicity.
This model mimics the spectral problems of a coupled system between even and odd gravitational modes.
The characteristic equation~\eqref{eq:matrixeigenvalueeq} for the matrix~\eqref{eq:matrixdeg} becomes
\begin{align}
    (2-\lambda)^2-\delta^2 =0.
    \label{eq: degenerate eigenequation}
\end{align}
Substituting $\lambda=2+\Delta\lambda$ into Eq.~\eqref{eq: degenerate eigenequation}, we obtain $(\Delta \lambda)^2 - \delta^2 = 0$.
This implies that the eigenvalues are  $\lambda = 2 \pm\delta$.
For symmetric matrices that are degenerate at the lowest order, the correction of the eigenvalue appears at ${\cal O}(\delta)$.
This is because the leading-correction term of $\Delta \lambda$ in the characteristic equation is ${\cal O}((\Delta \lambda)^2)$ if the lowest-order spectra are degenerate.

\subsection{$3 \times 3$  case}

Let us consider the eigenvalues of a $3 \times 3$ symmetric matrix, with a spectral feature similar to the coupled system among scalar, even, and odd gravitational modes calculated in Sec.~\ref{sec: QNM}.
We set a matrix ${\bf M}$ as
\begin{align}
    {\bf M} = \begin{pmatrix}
    1 & 0 & 0 \\
    0 &  2 & 0 \\
    0 & 0 & 2
    \end{pmatrix}
    +
    \begin{pmatrix}
    v_{\rm s}\, \epsilon^2 & b_2 \epsilon  & b_1\,  \epsilon \\
    b_2\, \epsilon &  v_{-}\, \epsilon^2 & b_3\, \epsilon^2 \\
    b_1\, \epsilon & b_3\, \epsilon^2  & v_{+}\, \epsilon^2
    \end{pmatrix},
    \label{eq:M_toy}
\end{align}
where $\epsilon$ is a small parameter, and the coefficients in front of $\epsilon$ and $\epsilon^2$ are constants.
As shown below, this choice of the correction terms
results in ${\cal O}(\epsilon^2)$ corrections to the eigenvalues of the matrix.

Substituting $\lambda = 1+f_1\epsilon^2 + {\cal O}(\epsilon^3)$ into the characteristic equation~\eqref{eq:matrixeigenvalueeq} for the matrix~\eqref{eq:M_toy},
we obtain
\begin{align}
    (v_{\rm s} -b_1^2 -b_2^2 -f_1)\epsilon^2 + {\cal O}(\epsilon^4)= 0.
\end{align}
Thus, the corresponding eigenvalue up to ${\cal O}(\epsilon^2)$ is
\begin{align}
\lambda = 1+(v_{\rm s} -b_1^2 -b_2^2)\epsilon^2.
\label{eq:lambdand}
\end{align}
Similar to the case of a $2 \times 2$ matrix with non-degenerate spectra at the lowest order,
squared terms of the non-diagonal components appear in the leading correction term in Eq.~\eqref{eq:lambdand}.

On the other hand, substituting $\lambda = 2 +f_2\epsilon^2 + {\cal O}(\epsilon^3)$ into the characteristic equation~\eqref{eq:matrixeigenvalueeq} for the matrix~\eqref{eq:M_toy},
we obtain
\begin{align}
    \{2b_1b_2b_3 +b_3^2 
    -b_1^2 v_- - b_2^2 v_+
    +
    (b_1^2 +b_2^2)f_2  -(f_2 -v_{-})(f_2 -v_{+})\}\epsilon^4 + {\cal O}(\epsilon^6)= 0.
\end{align}
Solving this equation at ${\cal O}(\epsilon^4)$,
the corresponding eigenvalues up to ${\cal O}(\epsilon^2)$ are determined as
\begin{align}
    \lambda = \lambda_{\rm I,II} = 2 +\frac{\epsilon^2}{2}\left(v_{-} +v_{+} + b_1^2 +b_2^2  \pm \sqrt{
    \left(v_- -v_+ + b_2^2 - b_1^2\right)^2 + 4(b_1 b_2 + b_3)^2
        }\right).
        \label{eq:evlambda}
\end{align}
Due to the degeneracy at the lowest order,
the spectra are expressed in a complicated form.
If $b_3 \propto b_1 b_2$ and $v_\pm \propto b_1^2$, the form of Eq.~\eqref{eq:evlambda}
is similar to the tensor-led quarinormal frequencies in Eqs.~\eqref{eq: omegaGrav correction I} and \eqref{eq: omegaGrav correction II}.

As an interesting limit, we study the case with $b_1 b_2 + b_3 = 0$ in Eq.~\eqref{eq:evlambda}.
In that case, the square root in Eq.~\eqref{eq:evlambda} becomes 
\begin{align}
\sqrt{\left(v_- -v_+ + b_2^2 - b_1^2\right)^2}
=
(-1)^p(v_- -v_+ + b_2^2 - b_1^2),
\end{align}
where $p = 0$ for ${\rm Re}[v_- -v_+ + b_2^2 - b_1^2] > 0$
and $p = 1$ for ${\rm Re}[v_- -v_+ + b_2^2 - b_1^2] < 0$.
In that case, the eigenvalues become
\begin{align}
& \lambda_{\rm I} = 2 +\epsilon^2
\left(v_{-} +b_2^2 \right),
   \quad
\lambda_{\rm II} = 2 +\epsilon^2
\left(v_{+} +b_1^2 \right),
\quad {\text{for}~} p=0,
\label{eq:p0branch}
\\
& \lambda_{\rm I} = 2 +\epsilon^2
\left(v_{+} +b_1^2 \right),
   \quad
\lambda_{\rm II} = 2 +\epsilon^2
\left(v_{-} +b_2^2 \right),
\quad {\text{for}~} p=1.
\label{eq:p1branch}
\end{align}
If $p$ takes different values when $b_1 =0$ and when $b_2 = 0$ while keeping the condition $b_1 b_2 + b_3 = 0$,
the branches of Eqs.~\eqref{eq:p0branch} and~\eqref{eq:p1branch} 
exchange
for $b_1 =0$ and $b_2 = 0$ cases.
This phenomenon can be seen in the QNM spectra for our setup.

\section{Static solutions as tidal response for $\ell=2$}\label{sec: full response}

In Sec.~\ref{sec: tidal}, we discussed the tidal response from the new even-odd couplings.
In this section, we present explicit forms of the static solutions 
with the regularity condition at the horizon
for $\ell = 2, \omega = 0$ in the master equations \eqref{eq: coupled master eq}.
To obtain perturbative solutions in $\epsilon$, we expand the solutions around the lowest order solutions up to ${\cal O}(\epsilon^2)$ as Eq.~\eqref{eq:staticpsi}.
The lowest-order solutions $\tilde{\Psi}^{\rm s}_{\rm Sch},~\tilde{\Psi}^-_{\rm Sch}$, and $\tilde{\Psi}^+_{\rm Sch}$ 
are given by Eqs.~\eqref{eq: 0th growing scalar}--\eqref{eq: 0th growing even}, 
which corresponds to the Schwarzschild case.
We derive the correction terms
by solving Eq.~\eqref{eq: coupled master eq} with the ansatz in Eq.~\eqref{eq:staticpsi}.

\subsection{Scalar field-led case}

First, we consider solutions led by the scalar external tidal field.
The lowest-order growing solution for the scalar field~\eqref{eq: 0th growing scalar} is given by
\begin{align}
\tilde{\Psi}^{\rm s}_{\rm Sch} &=C^{\rm s} \left(6 \frac{r^3}{r_{\rm H}^3}-6 \frac{r^2}{{r_{\rm H}^2}} +\frac{r}{r_{\rm H}}\right).
\end{align}
Through the scalar-odd coupling $V^{\rm s-}_{\rm 1st}$ and scalar-even coupling $V^{\rm s+}_{\rm 1st}$ and the self-scalar coupling $V^{\rm s}_{\rm 2nd}$ in Eq.~\eqref{eq: potential_correction}, the higher-order solutions are given by
\begin{align}
    \tilde{\Psi}^{\rm s}_{\rm 1st} &= 0,
\\
    \tilde{\Psi}^-_{\rm 1st} &= -
    \frac{1}{\sqrt{3} r r_H^3}2 b_2 C^{\rm s} \bigg\{
    144 r^4 \text{Li}_2\left(1-\frac{r}{r_{\rm H}}\right) +12 r_{\rm H} (6 r^2 r_{\rm H}+12 r^3+4 r r_{\rm H}^2+3 r_{\rm H}^3) \ln \left(\frac{r_{\rm H}}{r}\right)
\notag\\ 
    &\quad
    +72 r^4 \ln ^2\left(\frac{r}{r_{\rm H}}\right)-36 r^2 r_{\rm H}^2-144 r^3 r_{\rm H}+24 \pi ^2 r^4+128 r r_{\rm H}^3-117 r_{\rm H}^4
    \bigg\},
\\
    \tilde{\Psi}^+_{\rm 1st} &=
    \frac{1}{\sqrt{3} r r_{\rm H}^3 (4 r+3 r_{\rm H})}4 b_1 C^{\rm s} \bigg\{-144 (-3 r^2 r_{\rm H}^3+6 r^4 r_{\rm H}+4 r^5) \text{Li}_2\left(1-\frac{r}{r_{\rm H}}\right)+1008 r^3 r_{\rm H}^2
\notag\\ 
    &\quad
    +24 r^2 \ln \left(\frac{r}{r_{\rm H}}\right) \left(2 r_{\rm H} (12 r^2+24 r r_{\rm H}+13 r_{\rm H}^2)-3 (6 r^2 r_{\rm H}+4 r^3-3 r_{\rm H}^3) \ln \left(\frac{r}{r_{\rm H}}\right)\right)
\\ 
    &\quad
    +8 (53+9 \pi ^2) r^2 r_{\rm H}^3-144 (\pi ^2-4) r^4 r_{\rm H}-96 \pi ^2 r^5-282 r r_{\rm H}^4+15 r_{\rm H}^5\bigg\},
\\
    \tilde{\Psi}^{\rm s}_{\rm 2nd} &= X_{\rm s}^{{\rm s}-{\rm led}},
\\
    \tilde{\Psi}^{-}_{\rm 2nd} &= 0,
\\
    \tilde{\Psi}^{+}_{\rm 2nd} &= 0,
\end{align}
where $\text{Li}_2$ is the second polylogarithm function
\begin{align}
    \text{Li}_2(x):= - \int_0^1  dt \frac{\ln(1-tx)}{t},
\end{align}
and $\zeta(n)$ is the Riemann zeta function.
The function $X_{\rm s}^{{\rm s}-{\rm led}}$ 
is defined as the solution of the master equation for $\tilde{\Psi}_{\rm s}^{(2)}$ with the regularity at the horizon.
Although it is difficult to write an explicit form of 
the function $X_{\rm s}^{{\rm s}-{\rm led}}$, we can discuss its asymptotic behavior at spatial infinity.
In the asymptotic region, the non-vanishing solutions behave as
\begin{align}
    \tilde{\Psi}^{-}_{\rm 1st}|_{r\to \infty} &= -
    b_2 C^{\rm s} \left(96 \sqrt{3}  -\frac{72 \sqrt{3} r_{\rm H}}{r} +\frac{96 \sqrt{3} r_{\rm H}^2}{25 r^2}
    +\frac{8 r_{\rm H}^3}{\sqrt{3} r^3}
    +\frac{96 \sqrt{3} r_{\rm H}^4}{49 r^4}
    \right)
\notag\\ 
    &\quad -b_2 C^{\rm s} \left(\frac{96 \sqrt{3} r_{\rm H}^2}{5 r^2}
    +\frac{16 \sqrt{3} r_{\rm H}^3}{r^3}
    +\frac{96 \sqrt{3} r_{\rm H}^4}{7 r^4}
    \right) \ln \left(\frac{r}{r_{\rm H}}\right),
\\
    \tilde{\Psi}^{+}_{\rm 1st}|_{r\to \infty} &= b_1 C^{\rm s} \left(48 \sqrt{3} 
    -\frac{30 \sqrt{3} r_{\rm H}}{r}
    +\frac{1891 \sqrt{3} r_{\rm H}^2}{50 r^2}
    -\frac{17531 r_{\rm H}^3}{200 \sqrt{3} r^3}
    +\frac{744619 \sqrt{3} r_{\rm H}^4}{39200 r^4}
    \right)
\notag\\ 
    &\quad +b_1 C^{\rm s} \left(-\frac{192 \sqrt{3} r_{\rm H}^2}{5 r^2}
    -\frac{64 \sqrt{3} r_{\rm H}^3}{5 r^3}
    -\frac{1044 \sqrt{3} r_{\rm H}^4}{35 r^4}
    \right) \ln \left(\frac{r}{r_{\rm H}}\right),
\\
    \tilde{\Psi}^{\rm s}_{\rm 2nd}|_{r\to \infty} &= b_1^2 C^{\rm s} \bigg(
    -\frac{588 r^2}{5 r_{\rm H}^2}
    +\frac{2092 r}{35r_{\rm H}}
    -\frac{2248 }{7}
    +\frac{2064 r_{\rm H}}{7 r}
    +\frac{4 r_{\rm H}^2 (322560 \zeta (3)-372959)}{175 r^2}
    +\cdots
    \bigg)
\notag\\ 
    &\quad +b_1^2 C^{\rm s} \left(
    -\frac{576 r_{\rm H}^2}{7 r^2}
    -\frac{864 r_{\rm H}^3}{7 r^3}
    +\cdots
    \right) \ln \left(\frac{r}{r_{\rm H}}\right)
\notag\\ 
    &\quad +b_2^2 C^{\rm s} \bigg(
    \frac{16 r_{\rm H}^2 (8640 \zeta (3)-13451)}{75 r^2}
    +\frac{8 r_{\rm H}^3 (8640 \zeta (3)-6251)}{25 r^3}
    +\cdots
    \bigg)
\notag\\ 
    &\quad +b_2^2 C^{\rm s} \left(
    -\frac{1152 r_{\rm H}^5}{5 r^5}
    -\frac{1792 r_{\rm H}^6}{5 r^6}
    +\cdots
    \right) \ln \left(\frac{r}{r_{\rm H}}\right).
\end{align}

\subsection{Odd-led case}

Let us consider solutions led by the odd parity external tidal field.
The lowest-order growing odd solution is given by Eq.~\eqref{eq: 0th growing odd}
\begin{align}
    \tilde{\Psi}^{-}_{\rm Sch
    } &= C^- \frac{r^3}{r_{\rm H}^3}.\label{eq: odd-led source}
\end{align}
Through the scalar-odd couplings $V^{\rm s-}_{\rm 1st}$, the first-order solutions are modified as
\begin{align}
    \tilde{\Psi}^{\rm s}_{\rm 1st} &=
    16 \sqrt{3} b_2 C^- \frac{r}{r_{\rm H}^3} \bigg\{6 (6 r^2-6 r r_{\rm H}+r_H^2) \text{Li}_2\left(1-\frac{r}{r_{\rm H}}\right)+3 (6 r^2-6 r r_{\rm H}+r_{\rm H}^2) \ln ^2\left(\frac{r}{r_{\rm H}}\right)
\notag\\ 
    &\quad +6 \pi ^2 r^2-6 (6+\pi ^2) r r_{\rm H}+18 r_H (2 r-r_{\rm H}) \ln \left(\frac{r_{\rm H}}{r}\right)+(27+\pi ^2) r_{\rm H}^2
    \bigg\},
\\
    \tilde{\Psi}^{-}_{\rm 1st} &= 0,
\\
    \tilde{\Psi}^{+}_{\rm 1st} &= 0.
\end{align}
At second order, through the self-odd coupling $V^{-}_{\rm 2nd}$ and the even-odd gravitational coupling $V^{-+}_{\rm 2nd}$ in Eq.~\eqref{eq: potential_correction}, the second-order solutions are   
\begin{align}
    \tilde{\Psi}^{\rm s}_{\rm 2nd} &= 0,
\\
    \tilde{\Psi}^-_{\rm 2nd} &= X_-^{{\rm odd}-{\rm led}},
\\
    \tilde{\Psi}^+_{\rm 2nd} &= X_+^{{\rm odd}-{\rm led}},
\end{align}
where the functions $X_-^{{\rm odd}-{\rm led}}$ and $X_+^{{\rm odd}-{\rm led}}$ 
are defined as the solutions of the master equations for $\tilde{\Psi}^-_{\rm 2nd}$ and $\tilde{\Psi}^+_{\rm 2nd}$ with the regularity at the horizon.
In the asymptotic region, non-vanishing solutions behave as
\begin{align}
    \tilde{\Psi}^{\rm s}_{\rm 1st}|_{r\to \infty} &= -
    b_2 C^- \left(16 \sqrt{3} 
    -\frac{4 \sqrt{3} r_{\rm H}}{r}
    -\frac{172 r_{\rm H}^2}{25 \sqrt{3} r^2}
    -\frac{26 \sqrt{3} r_{\rm H}^3}{25 r^3}
    + \cdots
    \right)
\notag\\ 
    &\quad - b_2 C^- \left(
    \frac{16 \sqrt{3} r_{\rm H}^2}{5 r^2}
    +
    \frac{24 \sqrt{3} r_{\rm H}^3}{5 r^3}
    +\cdots
    \right) \ln \left(\frac{r}{r_{\rm H}}\right),
\\
    \tilde{\Psi}^-_{\rm 2nd}|_{r\to \infty} &=
    b_1^2 C^- \left(-32 
    +\frac{40 r_{\rm H}}{r}
    +\frac{1608 r_{\rm H}^2}{5 r^2}
    -\frac{968 r_{\rm H}^3}{3 r^3} + \cdots \right)
\notag\\ 
    &\quad+b_2^2 C^- \left(
    \frac{6912 r_{\rm H}^2 (8 \zeta (3)-9)}{5 r^2}
    +
    \frac{384 r_{\rm H}^3 (24 \zeta (3)-29)}{r^3} + \cdots \right),\label{eq: odd-led odd}
\\
    \tilde{\Psi}^+_{\rm 2nd}|_{r\to \infty} &= 
    b_1 b_2 C^- \left(-\frac{1728}{7}
    +\frac{208 r_{\rm H}}{7 r}
    +\frac{6 r_{\rm H}^2 (129024 \zeta (3)-154439)}{35 r^2}
    + \cdots
    \right)
\notag\\ 
    &\quad +b_1 b_2 C^- \left(
    \frac{1536 r_{\rm H}^2}{7 r^2}
    +\frac{512 r_{\rm H}^3}{7 r^3}
    + \cdots
    \right) \ln \left(\frac{r}{r_{\rm H}}\right).
\end{align}
In the limits of the sGB gravity ($b_2=0$) or the dCS gravity ($b_1=0$), the odd mode~\eqref{eq: odd-led odd} does not include the logarithmic terms. 
In those cases, using the definition of the tidal Love numbers in~\cite{Cardoso:2017cfl},
we obtain
\begin{align}
    {\rm sGB~case:}&~ k^{\rm B}_2 = \frac{12032}{15}b_1^2\epsilon^2,
    \\
    {\rm dCS~case:}&~ k^{\rm B}_2 = 
    \frac{18432}{5}\{8\zeta(3) -9\}b_2^2\epsilon^2.
\end{align}
In the limit of the dCS gravity ($b_1=0$), our result coincides with that in~\cite{Cardoso:2017cfl}.

\subsection{Even-led case}

We study solutions led by the even parity external tidal field.
The lowest-order growing solution is given by Eq.~\eqref{eq: 0th growing even} as
\begin{align}
    \tilde{\Psi}^+_{\rm Sch} &=
    \frac{C^+}{{4 +3r_{\rm H}/r}} \left(-6 \frac{r^3}{r_{\rm H}^3}-4 \frac{r^2}{r_{\rm H}^2}+3\right).
\end{align}
Through the scalar-even coupling $V^{\rm s+}_{\rm 1st}$ in~\eqref{eq: potential_correction}, the first-order solutions are given by
\begin{align}
    \tilde{\Psi}^{\rm s}_{\rm 1st} &=
    \frac{1}{rr_{\rm H}^3 (4 r+3 r_{\rm H})}
    4 \sqrt{3} b_1 C^+ \bigg\{48 r^2 (4 r+3 r_{\rm H}) (6 r^2-6 r r_{\rm H}+r_{\rm H}^2) \text{Li}_2\left(1-\frac{r}{r_{\rm H}}\right)
\notag\\ 
    &\quad +608 r^2 r_{\rm H}^3 +8 \pi ^2 r^2 (4 r+3 r_{\rm H}) (6 r^2-6 r r_{\rm H}+r_{\rm H}^2) -1152 r^4 r_{\rm H}-24 r r_{\rm H}^4+5 r_{\rm H}^5
\notag\\ 
    &\quad +24 r^2 (4 r+3 r_{\rm H}) \ln \left(\frac{r}{r_{\rm H}}\right) 
    \Big[(6 r^2-6 r r_{\rm H}+r_{\rm H}^2) \ln \left(\frac{r}{r_{\rm H}}\right)+6 r_{\rm H} (r_{\rm H}-2 r)\Big]\bigg\},
\\
    \tilde{\Psi}^{-}_{\rm 1st} &= 0,
\\
    \tilde{\Psi}^{+}_{\rm 1st} &= 0.
\end{align}
Via the new even-odd gravitational coupling $V^{-+}_{\rm 2nd}$ in Eq.~\eqref{eq: potential_correction}, the second-order solutions are 
\begin{align}
    \tilde{\Psi}^{\rm s}_{\rm 2nd} &= 0,
\\
    \tilde{\Psi}^{-}_{\rm 2nd} &= X_-^{{\rm even}-{\rm led}},
\\
    \tilde{\Psi}^{+}_{\rm 2nd} &= X_+^{{\rm even}-{\rm led}},
\end{align}
where the functions $X_-^{{\rm even}-{\rm led}}$ and $X_+^{{\rm even}-{\rm led}}$ 
are defined as the solutions of the master equations for $\tilde{\Psi}^{-}_{\rm 2nd}$ and $\tilde{\Psi}^{+}_{\rm 2nd}$ with the regularity at the horizon.
The asymptotic forms of the non-vanishing solutions are given by
\begin{align}
    \tilde{\Psi}^{\rm s}_{\rm 1st}|_{r\to \infty} &=
    b_1 C^+ \left(-8 \sqrt{3}
    -\frac{2 \sqrt{3} r_{\rm H}}{r}
    -\frac{613 r_{\rm H}^2}{50 \sqrt{3} r^2}
    -\frac{491 \sqrt{3} r_{\rm H}^3}{200 r^3}
    +\cdots
    \right)
\notag\\ 
    &\quad +b_1 C^+ \left(
    \frac{32 \sqrt{3} r_{\rm H}^2}{5 r^2}
    + \frac{48 \sqrt{3} r_{\rm H}^3}{5 r^3}
    +\cdots
    \right) \ln \left(\frac{r}{r_{\rm H}}\right),
\\
    \tilde{\Psi}^{-}_{\rm 2nd}|_{r\to \infty} &= -b_1 b_2 C^+ \left(-\frac{1728}{7}
    -\frac{440 r_{\rm H}}{7 r}
    +\frac{6 r_{\rm H}^2 (129024 \zeta (3)-154199)}{35 r^2}
    +\cdots
    \right)
\notag\\ 
    &\quad -b_1 b_2 C^+ \left(\frac{1536 r_{\rm H}^2}{7 r^2} + \frac{1280 r_{\rm H}^3}{7 r^3} + \cdots \right) \ln \left(\frac{r}{r_{\rm H}}\right),
\\
    \tilde{\Psi}^{+}_{\rm 2nd}|_{r\to \infty} &=
    b_1^2 C^+ \bigg(-\frac{1509 r^2}{70r_{\rm H}^2}
    +\frac{2927 r}{140r_{\rm H}}
    -\frac{667333}{3360}
    -\frac{198489 r_{\rm H}}{1120 r}
    +\cdots \bigg)
\notag\\ 
    &\quad +b_1^2 C^+ \left(
    \frac{1920 r_{\rm H}^2}{7 r^2}
    +\frac{640 r_{\rm H}^3}{7 r^3}
    +\cdots
    \right) \ln \left(\frac{r}{r_{\rm H}}\right).
\end{align}

\bibliography{Refs}

\begin{thebibliography}{94}
\expandafter\ifx\csname natexlab\endcsname\relax\def\natexlab#1{#1}\fi
\expandafter\ifx\csname bibnamefont\endcsname\relax
  \def\bibnamefont#1{#1}\fi
\expandafter\ifx\csname bibfnamefont\endcsname\relax
  \def\bibfnamefont#1{#1}\fi
\expandafter\ifx\csname citenamefont\endcsname\relax
  \def\citenamefont#1{#1}\fi
\expandafter\ifx\csname url\endcsname\relax
  \def\url#1{\texttt{#1}}\fi
\expandafter\ifx\csname urlprefix\endcsname\relax\def\urlprefix{URL }\fi
\providecommand{\bibinfo}[2]{#2}
\providecommand{\eprint}[2][]{\url{#2}}

\bibitem[{\citenamefont{Abbott
  et~al.}(2016{\natexlab{a}})}]{LIGOScientific:2016aoc}
\bibinfo{author}{\bibfnamefont{B.~P.} \bibnamefont{Abbott}}
  \bibnamefont{et~al.} (\bibinfo{collaboration}{LIGO Scientific, Virgo}),
  \bibinfo{journal}{Phys. Rev. Lett.} \textbf{\bibinfo{volume}{116}},
  \bibinfo{pages}{061102} (\bibinfo{year}{2016}{\natexlab{a}}),
  \eprint{1602.03837}.

\bibitem[{\citenamefont{Abbott
  et~al.}(2016{\natexlab{b}})}]{LIGOScientific:2016vbw}
\bibinfo{author}{\bibfnamefont{B.~P.} \bibnamefont{Abbott}}
  \bibnamefont{et~al.} (\bibinfo{collaboration}{LIGO Scientific, Virgo}),
  \bibinfo{journal}{Phys. Rev. D} \textbf{\bibinfo{volume}{93}},
  \bibinfo{pages}{122003} (\bibinfo{year}{2016}{\natexlab{b}}),
  \eprint{1602.03839}.

\bibitem[{\citenamefont{Nakamura et~al.}(1987)\citenamefont{Nakamura, Oohara,
  and Kojima}}]{Nakamura:1987zz}
\bibinfo{author}{\bibfnamefont{T.}~\bibnamefont{Nakamura}},
  \bibinfo{author}{\bibfnamefont{K.}~\bibnamefont{Oohara}}, \bibnamefont{and}
  \bibinfo{author}{\bibfnamefont{Y.}~\bibnamefont{Kojima}},
  \bibinfo{journal}{Prog. Theor. Phys. Suppl.} \textbf{\bibinfo{volume}{90}},
  \bibinfo{pages}{1} (\bibinfo{year}{1987}).

\bibitem[{\citenamefont{Kokkotas and Schmidt}(1999)}]{Kokkotas:1999bd}
\bibinfo{author}{\bibfnamefont{K.~D.} \bibnamefont{Kokkotas}} \bibnamefont{and}
  \bibinfo{author}{\bibfnamefont{B.~G.} \bibnamefont{Schmidt}},
  \bibinfo{journal}{Living Rev. Rel.} \textbf{\bibinfo{volume}{2}},
  \bibinfo{pages}{2} (\bibinfo{year}{1999}), \eprint{gr-qc/9909058}.

\bibitem[{\citenamefont{Nollert}(1999)}]{Nollert:1999ji}
\bibinfo{author}{\bibfnamefont{H.-P.} \bibnamefont{Nollert}},
  \bibinfo{journal}{Class. Quant. Grav.} \textbf{\bibinfo{volume}{16}},
  \bibinfo{pages}{R159} (\bibinfo{year}{1999}).

\bibitem[{\citenamefont{Ferrari and Gualtieri}(2008)}]{Ferrari:2007dd}
\bibinfo{author}{\bibfnamefont{V.}~\bibnamefont{Ferrari}} \bibnamefont{and}
  \bibinfo{author}{\bibfnamefont{L.}~\bibnamefont{Gualtieri}},
  \bibinfo{journal}{Gen. Rel. Grav.} \textbf{\bibinfo{volume}{40}},
  \bibinfo{pages}{945} (\bibinfo{year}{2008}), \eprint{0709.0657}.

\bibitem[{\citenamefont{Berti et~al.}(2009)\citenamefont{Berti, Cardoso, and
  Starinets}}]{Berti:2009kk}
\bibinfo{author}{\bibfnamefont{E.}~\bibnamefont{Berti}},
  \bibinfo{author}{\bibfnamefont{V.}~\bibnamefont{Cardoso}}, \bibnamefont{and}
  \bibinfo{author}{\bibfnamefont{A.~O.} \bibnamefont{Starinets}},
  \bibinfo{journal}{Class. Quant. Grav.} \textbf{\bibinfo{volume}{26}},
  \bibinfo{pages}{163001} (\bibinfo{year}{2009}), \eprint{0905.2975}.

\bibitem[{\citenamefont{Konoplya and Zhidenko}(2011)}]{Konoplya:2011qq}
\bibinfo{author}{\bibfnamefont{R.~A.} \bibnamefont{Konoplya}} \bibnamefont{and}
  \bibinfo{author}{\bibfnamefont{A.}~\bibnamefont{Zhidenko}},
  \bibinfo{journal}{Rev. Mod. Phys.} \textbf{\bibinfo{volume}{83}},
  \bibinfo{pages}{793} (\bibinfo{year}{2011}), \eprint{1102.4014}.

\bibitem[{\citenamefont{Hatsuda and Kimura}(2021)}]{Hatsuda:2021gtn}
\bibinfo{author}{\bibfnamefont{Y.}~\bibnamefont{Hatsuda}} \bibnamefont{and}
  \bibinfo{author}{\bibfnamefont{M.}~\bibnamefont{Kimura}},
  \bibinfo{journal}{Universe} \textbf{\bibinfo{volume}{7}},
  \bibinfo{pages}{476} (\bibinfo{year}{2021}), \eprint{2111.15197}.

\bibitem[{\citenamefont{Riess et~al.}(1998)}]{SupernovaSearchTeam:1998fmf}
\bibinfo{author}{\bibfnamefont{A.~G.} \bibnamefont{Riess}} \bibnamefont{et~al.}
  (\bibinfo{collaboration}{Supernova Search Team}), \bibinfo{journal}{Astron.
  J.} \textbf{\bibinfo{volume}{116}}, \bibinfo{pages}{1009}
  (\bibinfo{year}{1998}), \eprint{astro-ph/9805201}.

\bibitem[{\citenamefont{Perlmutter
  et~al.}(1999)}]{SupernovaCosmologyProject:1998vns}
\bibinfo{author}{\bibfnamefont{S.}~\bibnamefont{Perlmutter}}
  \bibnamefont{et~al.} (\bibinfo{collaboration}{Supernova Cosmology Project}),
  \bibinfo{journal}{Astrophys. J.} \textbf{\bibinfo{volume}{517}},
  \bibinfo{pages}{565} (\bibinfo{year}{1999}), \eprint{astro-ph/9812133}.

\bibitem[{\citenamefont{Penrose}(1965)}]{Penrose:1964wq}
\bibinfo{author}{\bibfnamefont{R.}~\bibnamefont{Penrose}},
  \bibinfo{journal}{Phys. Rev. Lett.} \textbf{\bibinfo{volume}{14}},
  \bibinfo{pages}{57} (\bibinfo{year}{1965}).

\bibitem[{\citenamefont{Hawking and Penrose}(1970)}]{Hawking:1970zqf}
\bibinfo{author}{\bibfnamefont{S.~W.} \bibnamefont{Hawking}} \bibnamefont{and}
  \bibinfo{author}{\bibfnamefont{R.}~\bibnamefont{Penrose}},
  \bibinfo{journal}{Proc. Roy. Soc. Lond. A} \textbf{\bibinfo{volume}{314}},
  \bibinfo{pages}{529} (\bibinfo{year}{1970}).

\bibitem[{\citenamefont{Hawking}(1976)}]{Hawking:1976ra}
\bibinfo{author}{\bibfnamefont{S.~W.} \bibnamefont{Hawking}},
  \bibinfo{journal}{Phys. Rev. D} \textbf{\bibinfo{volume}{14}},
  \bibinfo{pages}{2460} (\bibinfo{year}{1976}).

\bibitem[{\citenamefont{Metsaev and Tseytlin}(1987)}]{Metsaev:1987zx}
\bibinfo{author}{\bibfnamefont{R.~R.} \bibnamefont{Metsaev}} \bibnamefont{and}
  \bibinfo{author}{\bibfnamefont{A.~A.} \bibnamefont{Tseytlin}},
  \bibinfo{journal}{Nucl. Phys. B} \textbf{\bibinfo{volume}{293}},
  \bibinfo{pages}{385} (\bibinfo{year}{1987}).

\bibitem[{\citenamefont{Kanti et~al.}(1996)\citenamefont{Kanti, Mavromatos,
  Rizos, Tamvakis, and Winstanley}}]{Kanti:1995vq}
\bibinfo{author}{\bibfnamefont{P.}~\bibnamefont{Kanti}},
  \bibinfo{author}{\bibfnamefont{N.~E.} \bibnamefont{Mavromatos}},
  \bibinfo{author}{\bibfnamefont{J.}~\bibnamefont{Rizos}},
  \bibinfo{author}{\bibfnamefont{K.}~\bibnamefont{Tamvakis}}, \bibnamefont{and}
  \bibinfo{author}{\bibfnamefont{E.}~\bibnamefont{Winstanley}},
  \bibinfo{journal}{Phys. Rev. D} \textbf{\bibinfo{volume}{54}},
  \bibinfo{pages}{5049} (\bibinfo{year}{1996}), \eprint{hep-th/9511071}.

\bibitem[{\citenamefont{Green and Schwarz}(1984)}]{Green:1984sg}
\bibinfo{author}{\bibfnamefont{M.~B.} \bibnamefont{Green}} \bibnamefont{and}
  \bibinfo{author}{\bibfnamefont{J.~H.} \bibnamefont{Schwarz}},
  \bibinfo{journal}{Phys. Lett. B} \textbf{\bibinfo{volume}{149}},
  \bibinfo{pages}{117} (\bibinfo{year}{1984}).

\bibitem[{\citenamefont{Antoniadis et~al.}(1992)\citenamefont{Antoniadis, Gava,
  and Narain}}]{Antoniadis:1992sa}
\bibinfo{author}{\bibfnamefont{I.}~\bibnamefont{Antoniadis}},
  \bibinfo{author}{\bibfnamefont{E.}~\bibnamefont{Gava}}, \bibnamefont{and}
  \bibinfo{author}{\bibfnamefont{K.~S.} \bibnamefont{Narain}},
  \bibinfo{journal}{Phys. Lett. B} \textbf{\bibinfo{volume}{283}},
  \bibinfo{pages}{209} (\bibinfo{year}{1992}), \eprint{hep-th/9203071}.

\bibitem[{\citenamefont{Lue et~al.}(1999)\citenamefont{Lue, Wang, and
  Kamionkowski}}]{Lue:1998mq}
\bibinfo{author}{\bibfnamefont{A.}~\bibnamefont{Lue}},
  \bibinfo{author}{\bibfnamefont{L.-M.} \bibnamefont{Wang}}, \bibnamefont{and}
  \bibinfo{author}{\bibfnamefont{M.}~\bibnamefont{Kamionkowski}},
  \bibinfo{journal}{Phys. Rev. Lett.} \textbf{\bibinfo{volume}{83}},
  \bibinfo{pages}{1506} (\bibinfo{year}{1999}), \eprint{astro-ph/9812088}.

\bibitem[{\citenamefont{Jackiw and Pi}(2003)}]{Jackiw:2003pm}
\bibinfo{author}{\bibfnamefont{R.}~\bibnamefont{Jackiw}} \bibnamefont{and}
  \bibinfo{author}{\bibfnamefont{S.~Y.} \bibnamefont{Pi}},
  \bibinfo{journal}{Phys. Rev. D} \textbf{\bibinfo{volume}{68}},
  \bibinfo{pages}{104012} (\bibinfo{year}{2003}), \eprint{gr-qc/0308071}.

\bibitem[{\citenamefont{Alexander and Yunes}(2009)}]{Alexander:2009tp}
\bibinfo{author}{\bibfnamefont{S.}~\bibnamefont{Alexander}} \bibnamefont{and}
  \bibinfo{author}{\bibfnamefont{N.}~\bibnamefont{Yunes}},
  \bibinfo{journal}{Phys. Rept.} \textbf{\bibinfo{volume}{480}},
  \bibinfo{pages}{1} (\bibinfo{year}{2009}), \eprint{0907.2562}.

\bibitem[{\citenamefont{Mignemi and Stewart}(1993)}]{Mignemi:1992nt}
\bibinfo{author}{\bibfnamefont{S.}~\bibnamefont{Mignemi}} \bibnamefont{and}
  \bibinfo{author}{\bibfnamefont{N.~R.} \bibnamefont{Stewart}},
  \bibinfo{journal}{Phys. Rev. D} \textbf{\bibinfo{volume}{47}},
  \bibinfo{pages}{5259} (\bibinfo{year}{1993}), \eprint{hep-th/9212146}.

\bibitem[{\citenamefont{Antoniadis et~al.}(1994)\citenamefont{Antoniadis,
  Rizos, and Tamvakis}}]{Antoniadis:1993jc}
\bibinfo{author}{\bibfnamefont{I.}~\bibnamefont{Antoniadis}},
  \bibinfo{author}{\bibfnamefont{J.}~\bibnamefont{Rizos}}, \bibnamefont{and}
  \bibinfo{author}{\bibfnamefont{K.}~\bibnamefont{Tamvakis}},
  \bibinfo{journal}{Nucl. Phys. B} \textbf{\bibinfo{volume}{415}},
  \bibinfo{pages}{497} (\bibinfo{year}{1994}), \eprint{hep-th/9305025}.

\bibitem[{\citenamefont{Mignemi}(1995)}]{Mignemi:1993ce}
\bibinfo{author}{\bibfnamefont{S.}~\bibnamefont{Mignemi}},
  \bibinfo{journal}{Phys. Rev. D} \textbf{\bibinfo{volume}{51}},
  \bibinfo{pages}{934} (\bibinfo{year}{1995}), \eprint{hep-th/9303102}.

\bibitem[{\citenamefont{Daniel and Jenks}(2024)}]{Daniel:2024lev}
\bibinfo{author}{\bibfnamefont{T.}~\bibnamefont{Daniel}} \bibnamefont{and}
  \bibinfo{author}{\bibfnamefont{L.}~\bibnamefont{Jenks}}
  (\bibinfo{year}{2024}), \eprint{2403.09373}.

\bibitem[{\citenamefont{Horndeski}(1974)}]{Horndeski:1974wa}
\bibinfo{author}{\bibfnamefont{G.~W.} \bibnamefont{Horndeski}},
  \bibinfo{journal}{Int. J. Theor. Phys.} \textbf{\bibinfo{volume}{10}},
  \bibinfo{pages}{363} (\bibinfo{year}{1974}).

\bibitem[{\citenamefont{Deffayet et~al.}(2011)\citenamefont{Deffayet, Gao,
  Steer, and Zahariade}}]{Deffayet:2011gz}
\bibinfo{author}{\bibfnamefont{C.}~\bibnamefont{Deffayet}},
  \bibinfo{author}{\bibfnamefont{X.}~\bibnamefont{Gao}},
  \bibinfo{author}{\bibfnamefont{D.~A.} \bibnamefont{Steer}}, \bibnamefont{and}
  \bibinfo{author}{\bibfnamefont{G.}~\bibnamefont{Zahariade}},
  \bibinfo{journal}{Phys. Rev. D} \textbf{\bibinfo{volume}{84}},
  \bibinfo{pages}{064039} (\bibinfo{year}{2011}), \eprint{1103.3260}.

\bibitem[{\citenamefont{Kobayashi et~al.}(2011)\citenamefont{Kobayashi,
  Yamaguchi, and Yokoyama}}]{Kobayashi:2011nu}
\bibinfo{author}{\bibfnamefont{T.}~\bibnamefont{Kobayashi}},
  \bibinfo{author}{\bibfnamefont{M.}~\bibnamefont{Yamaguchi}},
  \bibnamefont{and} \bibinfo{author}{\bibfnamefont{J.}~\bibnamefont{Yokoyama}},
  \bibinfo{journal}{Prog. Theor. Phys.} \textbf{\bibinfo{volume}{126}},
  \bibinfo{pages}{511} (\bibinfo{year}{2011}), \eprint{1105.5723}.

\bibitem[{\citenamefont{Motohashi and Suyama}(2012)}]{Motohashi:2011ds}
\bibinfo{author}{\bibfnamefont{H.}~\bibnamefont{Motohashi}} \bibnamefont{and}
  \bibinfo{author}{\bibfnamefont{T.}~\bibnamefont{Suyama}},
  \bibinfo{journal}{Phys. Rev. D} \textbf{\bibinfo{volume}{85}},
  \bibinfo{pages}{044054} (\bibinfo{year}{2012}), \eprint{1110.6241}.

\bibitem[{\citenamefont{Crisostomi et~al.}(2018)\citenamefont{Crisostomi, Noui,
  Charmousis, and Langlois}}]{Crisostomi:2017ugk}
\bibinfo{author}{\bibfnamefont{M.}~\bibnamefont{Crisostomi}},
  \bibinfo{author}{\bibfnamefont{K.}~\bibnamefont{Noui}},
  \bibinfo{author}{\bibfnamefont{C.}~\bibnamefont{Charmousis}},
  \bibnamefont{and} \bibinfo{author}{\bibfnamefont{D.}~\bibnamefont{Langlois}},
  \bibinfo{journal}{Phys. Rev. D} \textbf{\bibinfo{volume}{97}},
  \bibinfo{pages}{044034} (\bibinfo{year}{2018}), \eprint{1710.04531}.

\bibitem[{\citenamefont{Yunes and Stein}(2011)}]{Yunes:2011we}
\bibinfo{author}{\bibfnamefont{N.}~\bibnamefont{Yunes}} \bibnamefont{and}
  \bibinfo{author}{\bibfnamefont{L.~C.} \bibnamefont{Stein}},
  \bibinfo{journal}{Phys. Rev. D} \textbf{\bibinfo{volume}{83}},
  \bibinfo{pages}{104002} (\bibinfo{year}{2011}), \eprint{1101.2921}.

\bibitem[{\citenamefont{Bl\'azquez-Salcedo
  et~al.}(2016)\citenamefont{Bl\'azquez-Salcedo, Macedo, Cardoso, Ferrari,
  Gualtieri, Khoo, Kunz, and Pani}}]{Blazquez-Salcedo:2016enn}
\bibinfo{author}{\bibfnamefont{J.~L.} \bibnamefont{Bl\'azquez-Salcedo}},
  \bibinfo{author}{\bibfnamefont{C.~F.~B.} \bibnamefont{Macedo}},
  \bibinfo{author}{\bibfnamefont{V.}~\bibnamefont{Cardoso}},
  \bibinfo{author}{\bibfnamefont{V.}~\bibnamefont{Ferrari}},
  \bibinfo{author}{\bibfnamefont{L.}~\bibnamefont{Gualtieri}},
  \bibinfo{author}{\bibfnamefont{F.~S.} \bibnamefont{Khoo}},
  \bibinfo{author}{\bibfnamefont{J.}~\bibnamefont{Kunz}}, \bibnamefont{and}
  \bibinfo{author}{\bibfnamefont{P.}~\bibnamefont{Pani}},
  \bibinfo{journal}{Phys. Rev. D} \textbf{\bibinfo{volume}{94}},
  \bibinfo{pages}{104024} (\bibinfo{year}{2016}), \eprint{1609.01286}.

\bibitem[{\citenamefont{Bl\'azquez-Salcedo
  et~al.}(2017)\citenamefont{Bl\'azquez-Salcedo, Khoo, and
  Kunz}}]{Blazquez-Salcedo:2017txk}
\bibinfo{author}{\bibfnamefont{J.~L.} \bibnamefont{Bl\'azquez-Salcedo}},
  \bibinfo{author}{\bibfnamefont{F.~S.} \bibnamefont{Khoo}}, \bibnamefont{and}
  \bibinfo{author}{\bibfnamefont{J.}~\bibnamefont{Kunz}},
  \bibinfo{journal}{Phys. Rev. D} \textbf{\bibinfo{volume}{96}},
  \bibinfo{pages}{064008} (\bibinfo{year}{2017}), \eprint{1706.03262}.

\bibitem[{\citenamefont{Bl\'azquez-Salcedo
  et~al.}(2019)\citenamefont{Bl\'azquez-Salcedo, Altaha~Motahar, Doneva, Khoo,
  Kunz, Mojica, Staykov, and Yazadjiev}}]{Blazquez-Salcedo:2018pxo}
\bibinfo{author}{\bibfnamefont{J.~L.} \bibnamefont{Bl\'azquez-Salcedo}},
  \bibinfo{author}{\bibfnamefont{Z.}~\bibnamefont{Altaha~Motahar}},
  \bibinfo{author}{\bibfnamefont{D.~D.} \bibnamefont{Doneva}},
  \bibinfo{author}{\bibfnamefont{F.~S.} \bibnamefont{Khoo}},
  \bibinfo{author}{\bibfnamefont{J.}~\bibnamefont{Kunz}},
  \bibinfo{author}{\bibfnamefont{S.}~\bibnamefont{Mojica}},
  \bibinfo{author}{\bibfnamefont{K.~V.} \bibnamefont{Staykov}},
  \bibnamefont{and} \bibinfo{author}{\bibfnamefont{S.~S.}
  \bibnamefont{Yazadjiev}}, \bibinfo{journal}{Eur. Phys. J. Plus}
  \textbf{\bibinfo{volume}{134}}, \bibinfo{pages}{46} (\bibinfo{year}{2019}),
  \eprint{1810.09432}.

\bibitem[{\citenamefont{Konoplya et~al.}(2019)\citenamefont{Konoplya, Zinhailo,
  and Stuchl\'\i{}k}}]{Konoplya:2019hml}
\bibinfo{author}{\bibfnamefont{R.~A.} \bibnamefont{Konoplya}},
  \bibinfo{author}{\bibfnamefont{A.~F.} \bibnamefont{Zinhailo}},
  \bibnamefont{and}
  \bibinfo{author}{\bibfnamefont{Z.}~\bibnamefont{Stuchl\'\i{}k}},
  \bibinfo{journal}{Phys. Rev. D} \textbf{\bibinfo{volume}{99}},
  \bibinfo{pages}{124042} (\bibinfo{year}{2019}), \eprint{1903.03483}.

\bibitem[{\citenamefont{Zinhailo}(2019)}]{Zinhailo:2019rwd}
\bibinfo{author}{\bibfnamefont{A.~F.} \bibnamefont{Zinhailo}},
  \bibinfo{journal}{Eur. Phys. J. C} \textbf{\bibinfo{volume}{79}},
  \bibinfo{pages}{912} (\bibinfo{year}{2019}), \eprint{1909.12664}.

\bibitem[{\citenamefont{Churilova and Stuchlik}(2020)}]{Churilova:2019sah}
\bibinfo{author}{\bibfnamefont{M.~S.} \bibnamefont{Churilova}}
  \bibnamefont{and} \bibinfo{author}{\bibfnamefont{Z.}~\bibnamefont{Stuchlik}},
  \bibinfo{journal}{Annals Phys.} \textbf{\bibinfo{volume}{418}},
  \bibinfo{pages}{168181} (\bibinfo{year}{2020}), \eprint{1910.12660}.

\bibitem[{\citenamefont{Bl\'azquez-Salcedo
  et~al.}(2020)\citenamefont{Bl\'azquez-Salcedo, Doneva, Kahlen, Kunz, Nedkova,
  and Yazadjiev}}]{Blazquez-Salcedo:2020caw}
\bibinfo{author}{\bibfnamefont{J.~L.} \bibnamefont{Bl\'azquez-Salcedo}},
  \bibinfo{author}{\bibfnamefont{D.~D.} \bibnamefont{Doneva}},
  \bibinfo{author}{\bibfnamefont{S.}~\bibnamefont{Kahlen}},
  \bibinfo{author}{\bibfnamefont{J.}~\bibnamefont{Kunz}},
  \bibinfo{author}{\bibfnamefont{P.}~\bibnamefont{Nedkova}}, \bibnamefont{and}
  \bibinfo{author}{\bibfnamefont{S.~S.} \bibnamefont{Yazadjiev}},
  \bibinfo{journal}{Phys. Rev. D} \textbf{\bibinfo{volume}{102}},
  \bibinfo{pages}{024086} (\bibinfo{year}{2020}), \eprint{2006.06006}.

\bibitem[{\citenamefont{Pierini and Gualtieri}(2021)}]{Pierini:2021jxd}
\bibinfo{author}{\bibfnamefont{L.}~\bibnamefont{Pierini}} \bibnamefont{and}
  \bibinfo{author}{\bibfnamefont{L.}~\bibnamefont{Gualtieri}},
  \bibinfo{journal}{Phys. Rev. D} \textbf{\bibinfo{volume}{103}},
  \bibinfo{pages}{124017} (\bibinfo{year}{2021}), \eprint{2103.09870}.

\bibitem[{\citenamefont{Pierini and Gualtieri}(2022)}]{Pierini:2022eim}
\bibinfo{author}{\bibfnamefont{L.}~\bibnamefont{Pierini}} \bibnamefont{and}
  \bibinfo{author}{\bibfnamefont{L.}~\bibnamefont{Gualtieri}},
  \bibinfo{journal}{Phys. Rev. D} \textbf{\bibinfo{volume}{106}},
  \bibinfo{pages}{104009} (\bibinfo{year}{2022}), \eprint{2207.11267}.

\bibitem[{\citenamefont{Yunes and Sopuerta}(2008)}]{Yunes:2007ss}
\bibinfo{author}{\bibfnamefont{N.}~\bibnamefont{Yunes}} \bibnamefont{and}
  \bibinfo{author}{\bibfnamefont{C.~F.} \bibnamefont{Sopuerta}},
  \bibinfo{journal}{Phys. Rev. D} \textbf{\bibinfo{volume}{77}},
  \bibinfo{pages}{064007} (\bibinfo{year}{2008}), \eprint{0712.1028}.

\bibitem[{\citenamefont{Cardoso and Gualtieri}(2009)}]{Cardoso:2009pk}
\bibinfo{author}{\bibfnamefont{V.}~\bibnamefont{Cardoso}} \bibnamefont{and}
  \bibinfo{author}{\bibfnamefont{L.}~\bibnamefont{Gualtieri}},
  \bibinfo{journal}{Phys. Rev. D} \textbf{\bibinfo{volume}{80}},
  \bibinfo{pages}{064008} (\bibinfo{year}{2009}), \bibinfo{note}{[Erratum:
  Phys.Rev.D 81, 089903 (2010)]}, \eprint{0907.5008}.

\bibitem[{\citenamefont{Molina et~al.}(2010)\citenamefont{Molina, Pani,
  Cardoso, and Gualtieri}}]{Molina:2010fb}
\bibinfo{author}{\bibfnamefont{C.}~\bibnamefont{Molina}},
  \bibinfo{author}{\bibfnamefont{P.}~\bibnamefont{Pani}},
  \bibinfo{author}{\bibfnamefont{V.}~\bibnamefont{Cardoso}}, \bibnamefont{and}
  \bibinfo{author}{\bibfnamefont{L.}~\bibnamefont{Gualtieri}},
  \bibinfo{journal}{Phys. Rev. D} \textbf{\bibinfo{volume}{81}},
  \bibinfo{pages}{124021} (\bibinfo{year}{2010}), \eprint{1004.4007}.

\bibitem[{\citenamefont{Srivastava et~al.}(2021)\citenamefont{Srivastava, Chen,
  and Shankaranarayanan}}]{Srivastava:2021imr}
\bibinfo{author}{\bibfnamefont{M.}~\bibnamefont{Srivastava}},
  \bibinfo{author}{\bibfnamefont{Y.}~\bibnamefont{Chen}}, \bibnamefont{and}
  \bibinfo{author}{\bibfnamefont{S.}~\bibnamefont{Shankaranarayanan}},
  \bibinfo{journal}{Phys. Rev. D} \textbf{\bibinfo{volume}{104}},
  \bibinfo{pages}{064034} (\bibinfo{year}{2021}), \eprint{2106.06209}.

\bibitem[{\citenamefont{Wagle et~al.}(2022)\citenamefont{Wagle, Yunes, and
  Silva}}]{Wagle:2021tam}
\bibinfo{author}{\bibfnamefont{P.}~\bibnamefont{Wagle}},
  \bibinfo{author}{\bibfnamefont{N.}~\bibnamefont{Yunes}}, \bibnamefont{and}
  \bibinfo{author}{\bibfnamefont{H.~O.} \bibnamefont{Silva}},
  \bibinfo{journal}{Phys. Rev. D} \textbf{\bibinfo{volume}{105}},
  \bibinfo{pages}{124003} (\bibinfo{year}{2022}), \eprint{2103.09913}.

\bibitem[{\citenamefont{Bl\'azquez-Salcedo
  et~al.}(2023)\citenamefont{Bl\'azquez-Salcedo, Khoo, Kunz, and
  Gonz\'alez-Romero}}]{Blazquez-Salcedo:2023hwg}
\bibinfo{author}{\bibfnamefont{J.~L.} \bibnamefont{Bl\'azquez-Salcedo}},
  \bibinfo{author}{\bibfnamefont{F.~S.} \bibnamefont{Khoo}},
  \bibinfo{author}{\bibfnamefont{J.}~\bibnamefont{Kunz}}, \bibnamefont{and}
  \bibinfo{author}{\bibfnamefont{L.~M.} \bibnamefont{Gonz\'alez-Romero}}
  (\bibinfo{year}{2023}), \eprint{2312.10754}.

\bibitem[{\citenamefont{Endlich et~al.}(2017)\citenamefont{Endlich, Gorbenko,
  Huang, and Senatore}}]{Endlich:2017tqa}
\bibinfo{author}{\bibfnamefont{S.}~\bibnamefont{Endlich}},
  \bibinfo{author}{\bibfnamefont{V.}~\bibnamefont{Gorbenko}},
  \bibinfo{author}{\bibfnamefont{J.}~\bibnamefont{Huang}}, \bibnamefont{and}
  \bibinfo{author}{\bibfnamefont{L.}~\bibnamefont{Senatore}},
  \bibinfo{journal}{JHEP} \textbf{\bibinfo{volume}{09}}, \bibinfo{pages}{122}
  (\bibinfo{year}{2017}), \eprint{1704.01590}.

\bibitem[{\citenamefont{Cardoso et~al.}(2018)\citenamefont{Cardoso, Kimura,
  Maselli, and Senatore}}]{Cardoso:2018ptl}
\bibinfo{author}{\bibfnamefont{V.}~\bibnamefont{Cardoso}},
  \bibinfo{author}{\bibfnamefont{M.}~\bibnamefont{Kimura}},
  \bibinfo{author}{\bibfnamefont{A.}~\bibnamefont{Maselli}}, \bibnamefont{and}
  \bibinfo{author}{\bibfnamefont{L.}~\bibnamefont{Senatore}},
  \bibinfo{journal}{Phys. Rev. Lett.} \textbf{\bibinfo{volume}{121}},
  \bibinfo{pages}{251105} (\bibinfo{year}{2018}), \eprint{1808.08962}.

\bibitem[{\citenamefont{Cano and Ruip\'erez}(2019)}]{Cano:2019ore}
\bibinfo{author}{\bibfnamefont{P.~A.} \bibnamefont{Cano}} \bibnamefont{and}
  \bibinfo{author}{\bibfnamefont{A.}~\bibnamefont{Ruip\'erez}},
  \bibinfo{journal}{JHEP} \textbf{\bibinfo{volume}{05}}, \bibinfo{pages}{189}
  (\bibinfo{year}{2019}), \bibinfo{note}{[Erratum: JHEP 03, 187 (2020)]},
  \eprint{1901.01315}.

\bibitem[{\citenamefont{de~Rham et~al.}(2020)\citenamefont{de~Rham, Francfort,
  and Zhang}}]{deRham:2020ejn}
\bibinfo{author}{\bibfnamefont{C.}~\bibnamefont{de~Rham}},
  \bibinfo{author}{\bibfnamefont{J.}~\bibnamefont{Francfort}},
  \bibnamefont{and} \bibinfo{author}{\bibfnamefont{J.}~\bibnamefont{Zhang}},
  \bibinfo{journal}{Phys. Rev. D} \textbf{\bibinfo{volume}{102}},
  \bibinfo{pages}{024079} (\bibinfo{year}{2020}), \eprint{2005.13923}.

\bibitem[{\citenamefont{Nomura and Yoshida}(2022)}]{Nomura:2021efi}
\bibinfo{author}{\bibfnamefont{K.}~\bibnamefont{Nomura}} \bibnamefont{and}
  \bibinfo{author}{\bibfnamefont{D.}~\bibnamefont{Yoshida}},
  \bibinfo{journal}{Phys. Rev. D} \textbf{\bibinfo{volume}{105}},
  \bibinfo{pages}{044006} (\bibinfo{year}{2022}), \eprint{2111.06273}.

\bibitem[{\citenamefont{Cano et~al.}(2022{\natexlab{a}})\citenamefont{Cano,
  Ganchev, Mayerson, and Ruip\'erez}}]{Cano:2022wwo}
\bibinfo{author}{\bibfnamefont{P.~A.} \bibnamefont{Cano}},
  \bibinfo{author}{\bibfnamefont{B.}~\bibnamefont{Ganchev}},
  \bibinfo{author}{\bibfnamefont{D.~R.} \bibnamefont{Mayerson}},
  \bibnamefont{and}
  \bibinfo{author}{\bibfnamefont{A.}~\bibnamefont{Ruip\'erez}},
  \bibinfo{journal}{JHEP} \textbf{\bibinfo{volume}{12}}, \bibinfo{pages}{120}
  (\bibinfo{year}{2022}{\natexlab{a}}), \eprint{2208.01044}.

\bibitem[{\citenamefont{Silva et~al.}(2023)\citenamefont{Silva, Ghosh, and
  Buonanno}}]{Silva:2022srr}
\bibinfo{author}{\bibfnamefont{H.~O.} \bibnamefont{Silva}},
  \bibinfo{author}{\bibfnamefont{A.}~\bibnamefont{Ghosh}}, \bibnamefont{and}
  \bibinfo{author}{\bibfnamefont{A.}~\bibnamefont{Buonanno}},
  \bibinfo{journal}{Phys. Rev. D} \textbf{\bibinfo{volume}{107}},
  \bibinfo{pages}{044030} (\bibinfo{year}{2023}), \eprint{2205.05132}.

\bibitem[{\citenamefont{Cano et~al.}(2023{\natexlab{a}})\citenamefont{Cano,
  Fransen, Hertog, and Maenaut}}]{Cano:2023jbk}
\bibinfo{author}{\bibfnamefont{P.~A.} \bibnamefont{Cano}},
  \bibinfo{author}{\bibfnamefont{K.}~\bibnamefont{Fransen}},
  \bibinfo{author}{\bibfnamefont{T.}~\bibnamefont{Hertog}}, \bibnamefont{and}
  \bibinfo{author}{\bibfnamefont{S.}~\bibnamefont{Maenaut}}
  (\bibinfo{year}{2023}{\natexlab{a}}), \eprint{2307.07431}.

\bibitem[{\citenamefont{Cayuso et~al.}(2023)\citenamefont{Cayuso, Figueras,
  Fran\c{c}a, and Lehner}}]{Cayuso:2023xbc}
\bibinfo{author}{\bibfnamefont{R.}~\bibnamefont{Cayuso}},
  \bibinfo{author}{\bibfnamefont{P.}~\bibnamefont{Figueras}},
  \bibinfo{author}{\bibfnamefont{T.}~\bibnamefont{Fran\c{c}a}},
  \bibnamefont{and} \bibinfo{author}{\bibfnamefont{L.}~\bibnamefont{Lehner}},
  \bibinfo{journal}{Phys. Rev. Lett.} \textbf{\bibinfo{volume}{131}},
  \bibinfo{pages}{111403} (\bibinfo{year}{2023}).

\bibitem[{\citenamefont{Cano et~al.}(2023{\natexlab{b}})\citenamefont{Cano,
  Fransen, Hertog, and Maenaut}}]{Cano:2023tmv}
\bibinfo{author}{\bibfnamefont{P.~A.} \bibnamefont{Cano}},
  \bibinfo{author}{\bibfnamefont{K.}~\bibnamefont{Fransen}},
  \bibinfo{author}{\bibfnamefont{T.}~\bibnamefont{Hertog}}, \bibnamefont{and}
  \bibinfo{author}{\bibfnamefont{S.}~\bibnamefont{Maenaut}},
  \bibinfo{journal}{Phys. Rev. D} \textbf{\bibinfo{volume}{108}},
  \bibinfo{pages}{024040} (\bibinfo{year}{2023}{\natexlab{b}}),
  \eprint{2304.02663}.

\bibitem[{\citenamefont{Melville}(2024)}]{Melville:2024zjq}
\bibinfo{author}{\bibfnamefont{S.}~\bibnamefont{Melville}}
  (\bibinfo{year}{2024}), \eprint{2401.05524}.

\bibitem[{\citenamefont{Cano et~al.}(2022{\natexlab{b}})\citenamefont{Cano,
  Fransen, Hertog, and Maenaut}}]{Cano:2021myl}
\bibinfo{author}{\bibfnamefont{P.~A.} \bibnamefont{Cano}},
  \bibinfo{author}{\bibfnamefont{K.}~\bibnamefont{Fransen}},
  \bibinfo{author}{\bibfnamefont{T.}~\bibnamefont{Hertog}}, \bibnamefont{and}
  \bibinfo{author}{\bibfnamefont{S.}~\bibnamefont{Maenaut}},
  \bibinfo{journal}{Phys. Rev. D} \textbf{\bibinfo{volume}{105}},
  \bibinfo{pages}{024064} (\bibinfo{year}{2022}{\natexlab{b}}),
  \eprint{2110.11378}.

\bibitem[{\citenamefont{Nakashi and Kimura}(2020)}]{Nakashi:2020phm}
\bibinfo{author}{\bibfnamefont{K.}~\bibnamefont{Nakashi}} \bibnamefont{and}
  \bibinfo{author}{\bibfnamefont{M.}~\bibnamefont{Kimura}},
  \bibinfo{journal}{Phys. Rev. D} \textbf{\bibinfo{volume}{102}},
  \bibinfo{pages}{084021} (\bibinfo{year}{2020}), \eprint{2008.04003}.

\bibitem[{\citenamefont{Satoh et~al.}(2008)\citenamefont{Satoh, Kanno, and
  Soda}}]{Satoh:2007gn}
\bibinfo{author}{\bibfnamefont{M.}~\bibnamefont{Satoh}},
  \bibinfo{author}{\bibfnamefont{S.}~\bibnamefont{Kanno}}, \bibnamefont{and}
  \bibinfo{author}{\bibfnamefont{J.}~\bibnamefont{Soda}},
  \bibinfo{journal}{Phys. Rev. D} \textbf{\bibinfo{volume}{77}},
  \bibinfo{pages}{023526} (\bibinfo{year}{2008}), \eprint{0706.3585}.

\bibitem[{\citenamefont{Bekenstein}(1972)}]{Bekenstein:1971hc}
\bibinfo{author}{\bibfnamefont{J.~D.} \bibnamefont{Bekenstein}},
  \bibinfo{journal}{Phys. Rev. D} \textbf{\bibinfo{volume}{5}},
  \bibinfo{pages}{1239} (\bibinfo{year}{1972}).

\bibitem[{\citenamefont{Bekenstein}(1995)}]{Bekenstein:1995un}
\bibinfo{author}{\bibfnamefont{J.~D.} \bibnamefont{Bekenstein}},
  \bibinfo{journal}{Phys. Rev. D} \textbf{\bibinfo{volume}{51}},
  \bibinfo{pages}{R6608} (\bibinfo{year}{1995}).

\bibitem[{\citenamefont{Shin'ichi et~al.}()\citenamefont{Shin'ichi, Kimura, and
  Yamaguchi}}]{ours}
\bibinfo{author}{\bibfnamefont{H.}~\bibnamefont{Shin'ichi}},
  \bibinfo{author}{\bibfnamefont{M.}~\bibnamefont{Kimura}}, \bibnamefont{and}
  \bibinfo{author}{\bibfnamefont{M.}~\bibnamefont{Yamaguchi}},
  \bibinfo{note}{in preparation}.

\bibitem[{\citenamefont{Minamitsuji and Maeda}(2023)}]{Minamitsuji:2023nvh}
\bibinfo{author}{\bibfnamefont{M.}~\bibnamefont{Minamitsuji}} \bibnamefont{and}
  \bibinfo{author}{\bibfnamefont{K.-i.} \bibnamefont{Maeda}},
  \bibinfo{journal}{Phys. Rev. D} \textbf{\bibinfo{volume}{108}},
  \bibinfo{pages}{084061} (\bibinfo{year}{2023}), \eprint{2308.01082}.

\bibitem[{\citenamefont{Regge and Wheeler}(1957)}]{Regge:1957td}
\bibinfo{author}{\bibfnamefont{T.}~\bibnamefont{Regge}} \bibnamefont{and}
  \bibinfo{author}{\bibfnamefont{J.~A.} \bibnamefont{Wheeler}},
  \bibinfo{journal}{Phys. Rev.} \textbf{\bibinfo{volume}{108}},
  \bibinfo{pages}{1063} (\bibinfo{year}{1957}).

\bibitem[{\citenamefont{Zerilli}(1970)}]{Zerilli:1970se}
\bibinfo{author}{\bibfnamefont{F.~J.} \bibnamefont{Zerilli}},
  \bibinfo{journal}{Phys. Rev. Lett.} \textbf{\bibinfo{volume}{24}},
  \bibinfo{pages}{737} (\bibinfo{year}{1970}).

\bibitem[{\citenamefont{Leaver}(1986)}]{Leaver:1986gd}
\bibinfo{author}{\bibfnamefont{E.~W.} \bibnamefont{Leaver}},
  \bibinfo{journal}{Phys. Rev. D} \textbf{\bibinfo{volume}{34}},
  \bibinfo{pages}{384} (\bibinfo{year}{1986}).

\bibitem[{\citenamefont{Nollert and Schmidt}(1992)}]{Nollert:1992ifk}
\bibinfo{author}{\bibfnamefont{H.-P.} \bibnamefont{Nollert}} \bibnamefont{and}
  \bibinfo{author}{\bibfnamefont{B.~G.} \bibnamefont{Schmidt}},
  \bibinfo{journal}{Phys. Rev. D} \textbf{\bibinfo{volume}{45}},
  \bibinfo{pages}{2617} (\bibinfo{year}{1992}).

\bibitem[{\citenamefont{Andersson}(1995)}]{Andersson:1995zk}
\bibinfo{author}{\bibfnamefont{N.}~\bibnamefont{Andersson}},
  \bibinfo{journal}{Phys. Rev. D} \textbf{\bibinfo{volume}{51}},
  \bibinfo{pages}{353} (\bibinfo{year}{1995}).

\bibitem[{\citenamefont{Andersson}(1997)}]{Andersson:1996cm}
\bibinfo{author}{\bibfnamefont{N.}~\bibnamefont{Andersson}},
  \bibinfo{journal}{Phys. Rev. D} \textbf{\bibinfo{volume}{55}},
  \bibinfo{pages}{468} (\bibinfo{year}{1997}), \eprint{gr-qc/9607064}.

\bibitem[{\citenamefont{Leaver}(1985)}]{Leaver:1985ax}
\bibinfo{author}{\bibfnamefont{E.~W.} \bibnamefont{Leaver}},
  \bibinfo{journal}{Proc. Roy. Soc. Lond. A} \textbf{\bibinfo{volume}{402}},
  \bibinfo{pages}{285} (\bibinfo{year}{1985}).

\bibitem[{\citenamefont{Chandrasekhar and
  Detweiler}(1975)}]{Chandrasekhar:1975zza}
\bibinfo{author}{\bibfnamefont{S.}~\bibnamefont{Chandrasekhar}}
  \bibnamefont{and} \bibinfo{author}{\bibfnamefont{S.~L.}
  \bibnamefont{Detweiler}}, \bibinfo{journal}{Proc. Roy. Soc. Lond. A}
  \textbf{\bibinfo{volume}{344}}, \bibinfo{pages}{441} (\bibinfo{year}{1975}).

\bibitem[{\citenamefont{Cardoso et~al.}(2019)\citenamefont{Cardoso, Kimura,
  Maselli, Berti, Macedo, and McManus}}]{Cardoso:2019mqo}
\bibinfo{author}{\bibfnamefont{V.}~\bibnamefont{Cardoso}},
  \bibinfo{author}{\bibfnamefont{M.}~\bibnamefont{Kimura}},
  \bibinfo{author}{\bibfnamefont{A.}~\bibnamefont{Maselli}},
  \bibinfo{author}{\bibfnamefont{E.}~\bibnamefont{Berti}},
  \bibinfo{author}{\bibfnamefont{C.~F.~B.} \bibnamefont{Macedo}},
  \bibnamefont{and} \bibinfo{author}{\bibfnamefont{R.}~\bibnamefont{McManus}},
  \bibinfo{journal}{Phys. Rev. D} \textbf{\bibinfo{volume}{99}},
  \bibinfo{pages}{104077} (\bibinfo{year}{2019}), \eprint{1901.01265}.

\bibitem[{\citenamefont{McManus et~al.}(2019)\citenamefont{McManus, Berti,
  Macedo, Kimura, Maselli, and Cardoso}}]{McManus:2019ulj}
\bibinfo{author}{\bibfnamefont{R.}~\bibnamefont{McManus}},
  \bibinfo{author}{\bibfnamefont{E.}~\bibnamefont{Berti}},
  \bibinfo{author}{\bibfnamefont{C.~F.~B.} \bibnamefont{Macedo}},
  \bibinfo{author}{\bibfnamefont{M.}~\bibnamefont{Kimura}},
  \bibinfo{author}{\bibfnamefont{A.}~\bibnamefont{Maselli}}, \bibnamefont{and}
  \bibinfo{author}{\bibfnamefont{V.}~\bibnamefont{Cardoso}},
  \bibinfo{journal}{Phys. Rev. D} \textbf{\bibinfo{volume}{100}},
  \bibinfo{pages}{044061} (\bibinfo{year}{2019}), \eprint{1906.05155}.

\bibitem[{\citenamefont{Kimura}(2020)}]{Kimura:2020mrh}
\bibinfo{author}{\bibfnamefont{M.}~\bibnamefont{Kimura}},
  \bibinfo{journal}{Phys. Rev. D} \textbf{\bibinfo{volume}{101}},
  \bibinfo{pages}{064031} (\bibinfo{year}{2020}), \eprint{2001.09613}.

\bibitem[{\citenamefont{Franchini and V\"olkel}(2023)}]{Franchini:2022axs}
\bibinfo{author}{\bibfnamefont{N.}~\bibnamefont{Franchini}} \bibnamefont{and}
  \bibinfo{author}{\bibfnamefont{S.~H.} \bibnamefont{V\"olkel}},
  \bibinfo{journal}{Phys. Rev. D} \textbf{\bibinfo{volume}{107}},
  \bibinfo{pages}{124063} (\bibinfo{year}{2023}), \eprint{2210.14020}.

\bibitem[{\citenamefont{Hatsuda and Kimura}(2024)}]{Hatsuda:2023geo}
\bibinfo{author}{\bibfnamefont{Y.}~\bibnamefont{Hatsuda}} \bibnamefont{and}
  \bibinfo{author}{\bibfnamefont{M.}~\bibnamefont{Kimura}},
  \bibinfo{journal}{Phys. Rev. D} \textbf{\bibinfo{volume}{109}},
  \bibinfo{pages}{044026} (\bibinfo{year}{2024}), \eprint{2307.16626}.

\bibitem[{\citenamefont{Cunha et~al.}(2018)\citenamefont{Cunha, Herdeiro, and
  Radu}}]{Cunha:2018uzc}
\bibinfo{author}{\bibfnamefont{P.~V.~P.} \bibnamefont{Cunha}},
  \bibinfo{author}{\bibfnamefont{C.~A.~R.} \bibnamefont{Herdeiro}},
  \bibnamefont{and} \bibinfo{author}{\bibfnamefont{E.}~\bibnamefont{Radu}},
  \bibinfo{journal}{Phys. Rev. D} \textbf{\bibinfo{volume}{98}},
  \bibinfo{pages}{104060} (\bibinfo{year}{2018}), \eprint{1808.06692}.

\bibitem[{\citenamefont{Tahara et~al.}(2023)\citenamefont{Tahara, Takahashi,
  Minamitsuji, and Motohashi}}]{Tahara:2023pyg}
\bibinfo{author}{\bibfnamefont{H.~W.~H.} \bibnamefont{Tahara}},
  \bibinfo{author}{\bibfnamefont{K.}~\bibnamefont{Takahashi}},
  \bibinfo{author}{\bibfnamefont{M.}~\bibnamefont{Minamitsuji}},
  \bibnamefont{and} \bibinfo{author}{\bibfnamefont{H.}~\bibnamefont{Motohashi}}
  (\bibinfo{year}{2023}), \eprint{2312.11899}.

\bibitem[{\citenamefont{Hinderer}(2008)}]{Hinderer:2007mb}
\bibinfo{author}{\bibfnamefont{T.}~\bibnamefont{Hinderer}},
  \bibinfo{journal}{Astrophys. J.} \textbf{\bibinfo{volume}{677}},
  \bibinfo{pages}{1216} (\bibinfo{year}{2008}), \bibinfo{note}{[Erratum:
  Astrophys.J. 697, 964 (2009)]}, \eprint{0711.2420}.

\bibitem[{\citenamefont{Binnington and Poisson}(2009)}]{Binnington:2009bb}
\bibinfo{author}{\bibfnamefont{T.}~\bibnamefont{Binnington}} \bibnamefont{and}
  \bibinfo{author}{\bibfnamefont{E.}~\bibnamefont{Poisson}},
  \bibinfo{journal}{Phys. Rev. D} \textbf{\bibinfo{volume}{80}},
  \bibinfo{pages}{084018} (\bibinfo{year}{2009}), \eprint{0906.1366}.

\bibitem[{\citenamefont{Damour and Nagar}(2009)}]{Damour:2009vw}
\bibinfo{author}{\bibfnamefont{T.}~\bibnamefont{Damour}} \bibnamefont{and}
  \bibinfo{author}{\bibfnamefont{A.}~\bibnamefont{Nagar}},
  \bibinfo{journal}{Phys. Rev. D} \textbf{\bibinfo{volume}{80}},
  \bibinfo{pages}{084035} (\bibinfo{year}{2009}), \eprint{0906.0096}.

\bibitem[{\citenamefont{E and C.M.Will}(2014)}]{Poisson:2014tb}
\bibinfo{author}{\bibfnamefont{P.}~\bibnamefont{E}} \bibnamefont{and}
  \bibinfo{author}{\bibnamefont{C.M.Will}}, \emph{\bibinfo{title}{Gravity:
  Newtonian, Post-Newtonian, Relativistic}} (\bibinfo{publisher}{Cambridge
  University Press}, \bibinfo{year}{2014}).

\bibitem[{\citenamefont{Cardoso et~al.}(2017)\citenamefont{Cardoso, Franzin,
  Maselli, Pani, and Raposo}}]{Cardoso:2017cfl}
\bibinfo{author}{\bibfnamefont{V.}~\bibnamefont{Cardoso}},
  \bibinfo{author}{\bibfnamefont{E.}~\bibnamefont{Franzin}},
  \bibinfo{author}{\bibfnamefont{A.}~\bibnamefont{Maselli}},
  \bibinfo{author}{\bibfnamefont{P.}~\bibnamefont{Pani}}, \bibnamefont{and}
  \bibinfo{author}{\bibfnamefont{G.}~\bibnamefont{Raposo}},
  \bibinfo{journal}{Phys. Rev. D} \textbf{\bibinfo{volume}{95}},
  \bibinfo{pages}{084014} (\bibinfo{year}{2017}), \bibinfo{note}{[Addendum:
  Phys.Rev.D 95, 089901 (2017)]}, \eprint{1701.01116}.

\bibitem[{\citenamefont{Kol and Smolkin}(2012)}]{Kol:2011vg}
\bibinfo{author}{\bibfnamefont{B.}~\bibnamefont{Kol}} \bibnamefont{and}
  \bibinfo{author}{\bibfnamefont{M.}~\bibnamefont{Smolkin}},
  \bibinfo{journal}{JHEP} \textbf{\bibinfo{volume}{02}}, \bibinfo{pages}{010}
  (\bibinfo{year}{2012}), \eprint{1110.3764}.

\bibitem[{\citenamefont{Hui et~al.}(2021)\citenamefont{Hui, Joyce, Penco,
  Santoni, and Solomon}}]{Hui:2020xxx}
\bibinfo{author}{\bibfnamefont{L.}~\bibnamefont{Hui}},
  \bibinfo{author}{\bibfnamefont{A.}~\bibnamefont{Joyce}},
  \bibinfo{author}{\bibfnamefont{R.}~\bibnamefont{Penco}},
  \bibinfo{author}{\bibfnamefont{L.}~\bibnamefont{Santoni}}, \bibnamefont{and}
  \bibinfo{author}{\bibfnamefont{A.~R.} \bibnamefont{Solomon}},
  \bibinfo{journal}{JCAP} \textbf{\bibinfo{volume}{04}}, \bibinfo{pages}{052}
  (\bibinfo{year}{2021}), \eprint{2010.00593}.

\bibitem[{\citenamefont{Ivanov and Zhou}(2023)}]{Ivanov:2022hlo}
\bibinfo{author}{\bibfnamefont{M.~M.} \bibnamefont{Ivanov}} \bibnamefont{and}
  \bibinfo{author}{\bibfnamefont{Z.}~\bibnamefont{Zhou}},
  \bibinfo{journal}{Phys. Rev. D} \textbf{\bibinfo{volume}{107}},
  \bibinfo{pages}{084030} (\bibinfo{year}{2023}), \eprint{2208.08459}.

\bibitem[{\citenamefont{Charalambous et~al.}(2022)\citenamefont{Charalambous,
  Dubovsky, and Ivanov}}]{Charalambous:2022rre}
\bibinfo{author}{\bibfnamefont{P.}~\bibnamefont{Charalambous}},
  \bibinfo{author}{\bibfnamefont{S.}~\bibnamefont{Dubovsky}}, \bibnamefont{and}
  \bibinfo{author}{\bibfnamefont{M.~M.} \bibnamefont{Ivanov}},
  \bibinfo{journal}{JHEP} \textbf{\bibinfo{volume}{10}}, \bibinfo{pages}{175}
  (\bibinfo{year}{2022}), \eprint{2209.02091}.

\bibitem[{\citenamefont{Saketh et~al.}(2023)\citenamefont{Saketh, Zhou, and
  Ivanov}}]{Saketh:2023bul}
\bibinfo{author}{\bibfnamefont{M.~V.~S.} \bibnamefont{Saketh}},
  \bibinfo{author}{\bibfnamefont{Z.}~\bibnamefont{Zhou}}, \bibnamefont{and}
  \bibinfo{author}{\bibfnamefont{M.~M.} \bibnamefont{Ivanov}}
  (\bibinfo{year}{2023}), \eprint{2307.10391}.

\bibitem[{\citenamefont{Mandal et~al.}(2024)\citenamefont{Mandal, Mastrolia,
  Silva, Patil, and Steinhoff}}]{Mandal:2023hqa}
\bibinfo{author}{\bibfnamefont{M.~K.} \bibnamefont{Mandal}},
  \bibinfo{author}{\bibfnamefont{P.}~\bibnamefont{Mastrolia}},
  \bibinfo{author}{\bibfnamefont{H.~O.} \bibnamefont{Silva}},
  \bibinfo{author}{\bibfnamefont{R.}~\bibnamefont{Patil}}, \bibnamefont{and}
  \bibinfo{author}{\bibfnamefont{J.}~\bibnamefont{Steinhoff}},
  \bibinfo{journal}{JHEP} \textbf{\bibinfo{volume}{02}}, \bibinfo{pages}{188}
  (\bibinfo{year}{2024}), \eprint{2308.01865}.

\bibitem[{\citenamefont{Creci et~al.}(2023)\citenamefont{Creci, Hinderer, and
  Steinhoff}}]{Creci:2023cfx}
\bibinfo{author}{\bibfnamefont{G.}~\bibnamefont{Creci}},
  \bibinfo{author}{\bibfnamefont{T.}~\bibnamefont{Hinderer}}, \bibnamefont{and}
  \bibinfo{author}{\bibfnamefont{J.}~\bibnamefont{Steinhoff}},
  \bibinfo{journal}{Phys. Rev. D} \textbf{\bibinfo{volume}{108}},
  \bibinfo{pages}{124073} (\bibinfo{year}{2023}), \eprint{2308.11323}.

\bibitem[{\citenamefont{De~Luca et~al.}(2023)\citenamefont{De~Luca, Khoury, and
  Wong}}]{DeLuca:2022tkm}
\bibinfo{author}{\bibfnamefont{V.}~\bibnamefont{De~Luca}},
  \bibinfo{author}{\bibfnamefont{J.}~\bibnamefont{Khoury}}, \bibnamefont{and}
  \bibinfo{author}{\bibfnamefont{S.~S.~C.} \bibnamefont{Wong}},
  \bibinfo{journal}{Phys. Rev. D} \textbf{\bibinfo{volume}{108}},
  \bibinfo{pages}{044066} (\bibinfo{year}{2023}), \eprint{2211.14325}.

\bibitem[{\citenamefont{Katagiri et~al.}(2024)\citenamefont{Katagiri, Ikeda,
  and Cardoso}}]{Katagiri:2023umb}
\bibinfo{author}{\bibfnamefont{T.}~\bibnamefont{Katagiri}},
  \bibinfo{author}{\bibfnamefont{T.}~\bibnamefont{Ikeda}}, \bibnamefont{and}
  \bibinfo{author}{\bibfnamefont{V.}~\bibnamefont{Cardoso}},
  \bibinfo{journal}{Phys. Rev. D} \textbf{\bibinfo{volume}{109}},
  \bibinfo{pages}{044067} (\bibinfo{year}{2024}), \eprint{2310.19705}.

\bibitem[{\citenamefont{Bhattacharyya et~al.}(2024)\citenamefont{Bhattacharyya,
  Ghosh, Ghosh, and Pal}}]{Bhattacharyya:2024aeq}
\bibinfo{author}{\bibfnamefont{A.}~\bibnamefont{Bhattacharyya}},
  \bibinfo{author}{\bibfnamefont{D.}~\bibnamefont{Ghosh}},
  \bibinfo{author}{\bibfnamefont{S.}~\bibnamefont{Ghosh}}, \bibnamefont{and}
  \bibinfo{author}{\bibfnamefont{S.}~\bibnamefont{Pal}},
  \bibinfo{journal}{JHEP} \textbf{\bibinfo{volume}{04}}, \bibinfo{pages}{015}
  (\bibinfo{year}{2024}), \eprint{2401.05492}.

\end{thebibliography}

\end{document}